\documentclass[10pt,journal,compsoc]{IEEEtran}
\IEEEoverridecommandlockouts
\usepackage{booktabs} 
\usepackage{pbox}
\usepackage{csvsimple}
\usepackage{amssymb}
\usepackage{mathtools}
\usepackage{rotating}
\usepackage{multicol}
\usepackage{multirow}
\usepackage{xcolor}
\usepackage[export]{adjustbox}
\usepackage{array}
\usepackage{tabularx}
\usepackage{boldline}
\usepackage{bigstrut}
\usepackage{graphicx}
\usepackage{pgfplots}
\usepackage{soul}
\usepackage{paralist}
\usepackage{enumitem}
\usepackage{amsmath}
\usepackage{hyperref}
\usepackage{cleveref}
\usepackage{parcolumns}
\usepackage{pgfplots}
\usepackage[breakable, theorems, skins]{tcolorbox}
\usepackage{marvosym}
\usepackage[flushleft]{threeparttable}
\pgfplotsset{compat=1.3}
\usepackage{color, colortbl}
\usepackage[utf8]{inputenc}
\usepackage[english]{babel}
\usepackage{pgf}
\usepackage{balance}
\usetikzlibrary{positioning}
\usepackage[linesnumbered,ruled,vlined]{algorithm2e}
\usetikzlibrary{arrows}
\usepackage{subcaption}
\usepackage{listings}
\usepackage[noadjust]{cite}
\usepackage{threeparttable}

\graphicspath{ {figures/} }

\usepackage{amssymb}
\usepackage{pifont}
\newcommand{\xmark}{\ding{55}}%

\newcommand{\motivation}{\noindent\textbf{Motivation}.~\xspace}
\newcommand{\approach}{\noindent\textbf{Approach}.~\xspace}
\newcommand{\observations}{\noindent\textbf{Observations}.~\xspace}
\newcommand{\vp}{vulnerability prediction\xspace}
\newcommand{\VP}{Vulnerability prediction\xspace}

\newcommand{\dlvp}{DLVP\xspace}
\newcommand{\devign}{Devign\xspace}

\newcommand{\realdata}{\tool dataset\xspace}

\newcommand{\devigndata}{FFMPeg+Qemu\xspace}

\newcommand{\Hide}[1]{}
\newcommand{\be}{\begin{enumerate}[wide=0pt]}
\newcommand{\ee}{\end{enumerate}}
\newcommand{\bi}{\begin{itemize}[wide=0pt]}
\newcommand{\ei}{\end{itemize}}
\newcommand{\tool}{\textsc{ReVeal}\xspace}
\newcommand{\draper}{Draper\xspace}

\usepackage[framemethod=tikz]{mdframed}
\usetikzlibrary{shadows}
\usepackage{graphics}
\newmdenv[
    tikzsetting= {fill=gray!8},
    skipabove=0.4em,
    skipbelow=0.4em,
    linewidth=1pt,
    innerleftmargin=3pt,
    innerrightmargin=3pt,
    innertopmargin=2pt,
    innerbottommargin=2pt,
    linecolor=gray80,
    roundcorner=3pt, 
    shadow=true,
    shadowsize=5pt,
    shadowcolor=gray80
]{myshadowbox}
\usepackage{tikz}

\newcolumntype{s}{>{\centering \arraybackslash \hsize=.5\hsize}X}  %

\newenvironment{result}
{\begin{myshadowbox}\textbf{Result:}~}
{\end{myshadowbox}}

\newcommand\ProcessOperators{%
  \ifnum\lst@mode=\lst@Pmode%
    \def\lst@thestyle{\opstyle}
  \fi%
}

\newcommand{\lone}[1]{\colorbox{red!30}{#1}}
\newcommand{\ltwo}[1]{\colorbox{orange!30}{#1}}
\newcommand{\lthree}[1]{\colorbox{orange!20}{#1}}
\newcommand{\lfour}[1]{\colorbox{green!20}{#1}}
\newcommand{\lfive}[1]{\colorbox{green!30}{#1}}
\newcommand{\lsix}[1]{\colorbox{green!95}{#1}}

\DeclareFontFamily{U}{stixscr}{}
\DeclareFontShape{U}{stixscr}{m}{n}{<-> s*[1.1] stix-mathscr}{}

\SetCommentSty{mycommfont}

\newcommand{\algoframework}{%
\begingroup
\removelatexerror 
\begin{algorithm*}[H]
    \SetAlgoLined
    \footnotesize
    \SetKwInOut{Input}{Input}
    \SetKwInOut{Output}{Output}
    \SetKwFunction{Fframework}{\tool}
    \SetKwProg{Fn}{Function}{:}{}
    \Input{Train data -- $\mathbb{D}_\text{train}$, \\
         Contribution of triplet loss -- $\alpha$, \\
         Contribution of regularization loss -- $\beta$, \\
         Separation boundary -- $\gamma$, \\
         Learning rate -- $lr$}
    \Output{Trained model.}
    \Fn{\Fframework}{
        $\mathit{features} \leftarrow \emptyset$
        
        $\mathit{labels} \leftarrow \emptyset$
        
        \smallskip$\triangleright$~\texttt{Extract features from every code}
        
        \For{$(\mathbb{C}, l) \in \mathbb{D}_\text{train}$}{
            $f \leftarrow $ embed\_features($\mathbb{C}$)\\
            $\mathit{features} \leftarrow \mathit{features} \cup f$\\
            $\mathit{labels} \leftarrow \mathit{labels} \cup l$\\
        }
        
        \smallskip$\triangleright$~\texttt{Rebalance with SMOTE.}
        
        $\mathbb{D}_{balanced} \leftarrow $SMOTE($\mathit{features},~\mathit{labels}$)\\
        $\mathbb{M} \leftarrow $ RepresentationLearningModel()\\
        
        \smallskip$\triangleright$~\texttt{Train Model.}\label{alg_line:training}

        \For{$(x_g, l_{x_g}) \in \mathbb{D}_{balanced}$}{
            \smallskip$\triangleright$~\texttt{Define loss function.}\\
            $\mathbb{L}_{all} \leftarrow  loss\_function(\mathbb{M},~\mathbb{D}_{balanced},~x_g,~l_{x_g},~\alpha,~\beta,~\gamma)$\\
            \smallskip$\triangleright$~\texttt{$\theta$ represents the model parameters of $\mathbb{M}$}.
            
            $\theta \leftarrow \theta -  \nabla_{\theta}(\mathbb{L}_{all})$\label{alg_line:opt}
        }
        \Return $\mathbb{M_{\theta}}$
    }
    \caption{\textbf{\small \tool.}}
    \label{alg:overview_psuedocode}
\end{algorithm*}
\endgroup}

\makeatletter
\newcommand{\removelatexerror}{\let\@latex@error\@gobble}
\makeatother

\newcommand{\algosmote}{
\begingroup
\removelatexerror
\begin{algorithm}[H]
    \SetAlgoLined
    \footnotesize
    \SetKwInOut{Input}{Input}
    \SetKwInOut{Output}{Output}
    \SetKwFunction{SMOTE}{SMOTE}
    \SetKwProg{Fn}{Function}{:}{}
    \Input{Training Dataset -- $\mathbb{D}_\text{train}$\\
           Number of Nearest Neighbors -- $k$, \\
           Expected Number of samples per class -- $m$
         }
    \Output{Sampled Dataset -- $\mathbb{D}_\text{sampled}$.}
    \Fn{\SMOTE{$\mathbb{D}_\text{train}$, $k$, $m$}}{
        $\mathbb{D}_\text{sampled} \leftarrow \mathbb{D}_\text{train}$\\
        \While{\# of Majority examples $> m$}{
            $x \leftarrow $ random majority class example from $\mathbb{D}_\text{sampled}$\\
            Remove $x$ from $\mathbb{D}_\text{sampled}$\\
        }
        \While{\# of Minority examples $< m$}{
            $x \leftarrow $ random minority class example from $\mathbb{D}_\text{sampled}$\\
            $neighbors \leftarrow k$ nearest minority neighbors of $x$\\
            \For{$n \in neighbors$}{
                $x_s \leftarrow interpolate(x, n)$\\
                Add $x_s$ to $\mathbb{D}_\text{sampled}$\\
            }
        }
        \Return $\mathbb{D}_\text{sampled}$\\
    }
    \caption{\textbf{\small Pseudocode for SMOTE.}}
    \label{alg:smote}
\end{algorithm}
\endgroup
}

\newcommand{\algofeature}{%
\begingroup
\removelatexerror 
\begin{algorithm*}[H]
    \footnotesize
    \SetAlgoLined
    \SetKwInOut{Input}{Input}
    \SetKwInOut{Output}{Output}
    \SetKwFunction{Ffeature}{embed\_{features}}
    \SetKwProg{Fn}{Function}{:}{}
    \Input{Code -- $\mathbb{C}$.}
    \Output{Feature vector $x_g$ representing $\mathbb{C}$.}
    \Fn{\Ffeature{$\mathbb{C}$}}{
        $(\mathbb{V}, \mathbb{E}) \leftarrow$ extract\_code\_property\_graph $(\mathbb{C})$\label{alg_line:emb_extract_graph}\\
        $\mathbb{X} \leftarrow \emptyset$\\
         \For{$v \in \mathbb{V}$}{
            $T_v \leftarrow onehot(v.type())$\label{alg_line:onehot}\\
            $C_v \leftarrow word2vec(v.code\_fragment())$\label{alg_line:word2vec}\\
            $x_v \leftarrow concat(T_v,~C_{v})$\label{alg_line:emb_graph_concat}\\
            $\mathbb{X} \leftarrow x_v \cup X$\\
         }
         $\mathbb{X}^\prime \leftarrow GGNN(\mathbb{X}, \mathbb{E})$\label{alg_line:emb_graph_ggnn}\\
         $x_g \leftarrow Aggregate(\mathbb{X}^\prime)$\label{alg_line:emb_graph_aggregate}\\
         \Return $x_g$
    }
    \caption{\textbf{\small Graph Embedding}}
    \label{alg:feature_extraction}
\end{algorithm*}
\endgroup}

\newcommand{\algolossfunc}{%
\begingroup
\removelatexerror 
\begin{algorithm*}[H]
    \footnotesize
    \SetAlgoLined
    \SetKwInOut{Input}{Input}
    \SetKwInOut{Output}{Output}
    \SetKwFunction{Floss}{Loss}
    \SetKwProg{Fn}{Function}{:}{}
    \Input{Model -- $\mathbb{M}$\\
           Sampled dataset -- $\mathbb{D}_{balanced}$\\
           Code feature and label of an example  -- $x_g, l_{x_g}$\\
           Training Hyperparameters -- $\alpha, \beta, \gamma, lr$
    }
    \Output{\tool loss.}
    \Fn{\Floss{$\mathbb{M}, \mathbb{D}_{balanced}, x_g, l_{x_g}, \alpha, \beta, \gamma, lr$}}{
        $\triangleright$~\texttt{Sample for $x_p$ and $x_n$ from $\mathbb{D}_{balanced}$.}\\
        $x_p \leftarrow x\in\mathbb{D}_{balanced} | (l_x = l_{x_g} \& x_p \neq x_g)$\\
        $x_n \leftarrow x\in\mathbb{D}_{balanced} | (l_x \neq l_{x_g} \& x_n \neq x_g)$\\
        \smallskip$\triangleright$~\texttt{Transform $x_g, x_p, x_n$ to latent space.}\\
        $h_g, y_g \leftarrow \mathbb{M}.predict(x_g)$\\
        $h_p \leftarrow \mathbb{M}.transform(x_p)$\\
        $h_n \leftarrow \mathbb{M}.transform(x_n)$\\
        $\mathbb{L}_{ce} \leftarrow $ cross\_entropy($y_g, l_{x_g}$)\\
        $\mathbb{L}_{dist} \leftarrow \left|\mathbb{D}(h_g, h_p) - \mathbb{D}(h_g, h_n) + \gamma \right|$\\
        $\mathbb{L}_{re} \leftarrow ||h(x_g)|| + ||h(x_p)|| + ||h(x_n)||$\\
        \smallskip$\triangleright$~\texttt{Final loss.}\\
        $\mathbb{L}_{all} \leftarrow \mathbb{L}_{ce} + \alpha*\mathbb{L}_{dist} + \beta*\mathbb{L}_{re}$\\
        \Return $\mathbb{L}_{all}$\\
    }
    \caption{\textbf{\small Loss function.}}
    \label{alg:loss_function}
\end{algorithm*}
\endgroup}

\definecolor{pink}{RGB}{252,145,149}
\definecolor{lightpink}{RGB}{252,145,149}
\definecolor{lightgray}{gray}{0.8}
\definecolor{darkgray}{gray}{0.6}
\definecolor{Gray}{rgb}{0.88,1,1}
\definecolor{Gray}{gray}{0.85}
\definecolor{Blue}{RGB}{0,29,193}
\definecolor{MyDarkBlue}{rgb}{0,0.08,0.45} 
\definecolor{pink}{RGB}{231,95,110}
\definecolor{lightergray}{rgb}{0.85, 0.85, 0.85}
\definecolor{lightestgray}{rgb}{0.95, 0.95, 0.95}
\definecolor{codebg}{HTML}{F4F4F4}

\definecolor{gray05}{gray}{0.95}
\definecolor{gray10}{gray}{0.90}
\definecolor{gray12}{gray}{0.88}
\definecolor{gray15}{gray}{0.85}
\definecolor{gray20}{gray}{0.80}
\definecolor{gray25}{gray}{0.75}
\definecolor{gray30}{gray}{0.70}
\definecolor{gray40}{gray}{0.60}
\definecolor{gray50}{gray}{0.50}
\definecolor{gray60}{gray}{0.40}
\definecolor{gray70}{gray}{0.30}
\definecolor{gray75}{gray}{0.25}
\definecolor{gray80}{gray}{0.20}
\definecolor{gray90}{gray}{0.10}

\everymath{\displaystyle}

\newcommand{\rom}[1]{\uppercase\expandafter{\romannumeral #1\relax}}

\newcommand{\etal}{\hbox{\emph{et al.}}\xspace}
\newcommand{\eg}{\hbox{\emph{e.g.,}}\xspace}
\newcommand{\ie}{\hbox{\emph{i.e.}}\xspace}
\newcommand{\wrt}{\hbox{\emph{w.r.t.}}\xspace}

\definecolor{gray50}{gray}{.5}
\definecolor{gray40}{gray}{.6}
\definecolor{gray30}{gray}{.7}
\definecolor{gray20}{gray}{.8}
\definecolor{gray10}{gray}{.9}
\definecolor{gray05}{gray}{.95}

\newenvironment{examplebox}{\par\begingroup
   \setlength{\fboxsep}{5pt}\findlength
   \setbox0=\vbox\bgroup\noindent
   \hsize=0.95\linewidth
   \begin{minipage}{0.95\linewidth}\normalsize}
    {\end{minipage}\egroup
    \textcolor{gray20}{\fboxsep1.5pt\fbox
     {\fboxsep5pt\colorbox{gray05}{\normalcolor\box0}}}
    \endgroup\par\noindent
    \normalcolor\ignorespacesafterend}

\newcounter{RQCounter}
\newcounter{RQACounter}

\definecolor{javared}{rgb}{0.6,0,0} 
\definecolor{javagreen}{rgb}{0.25,0.5,0.35} 
\definecolor{javapurple}{rgb}{0.5,0,0.35} 
\definecolor{javadocblue}{rgb}{0.25,0.35,0.75} 

\lstdefinestyle{customc}{
  belowcaptionskip=\baselineskip,
  breaklines=true,
  xleftmargin=\parindent,
  language=java,
  showstringspaces=false,
  basicstyle=\scriptsize\ttfamily,
  keywordstyle=\bfseries\color{javapurple},
  commentstyle=\itshape\blue,
}

\usepackage{color}

\definecolor{dkgreen}{rgb}{0,0.6,0}
\definecolor{gray}{rgb}{0.5,0.5,0.5}
\definecolor{mauve}{rgb}{0.58,0,0.82}

\definecolor{shadecolor}{RGB}{150,150,150}

\newcommand\Red[1]{\textcolor[rgb]{1.00,0.00,0.00}{\textbf{#1}}}
\newcommand\red[1]{\textcolor[rgb]{1.00,0.00,0.00}{#1}}

\newcommand\blue[1]{\textcolor[rgb]{0.00,0.00,1.00}{{#1}}}

\newcommand\green[1]{\textcolor[rgb]{0.0,0.6,0}{#1}}

\usepackage{pifont}

\newcommand{\fig}[1]{Figure~\ref{fig:#1}}

\newcommand{\tab}[1]{Table~\ref{tab:#1}}
\newcommand{\tion}[1]{\S\ref{sect:#1}}

\newcommand{\squishlist}{
 \begin{list}{$\circ$}
 { \setlength{\itemsep}{0pt}
   \setlength{\parsep}{1pt}
   \setlength{\topsep}{1pt}
   \setlength{\partopsep}{0pt}
   \setlength{\leftmargin}{1.5em}
   \setlength{\labelwidth}{1em}
   \setlength{\labelsep}{0.5em} } }

\newcommand{\squishlisttwo}{
 \begin{list}{$\bullet$}
 { \setlength{\itemsep}{0pt}
  \setlength{\parsep}{0pt}
  \setlength{\topsep}{0pt}
  \setlength{\partopsep}{0pt}
  \setlength{\leftmargin}{0em}
  \setlength{\labelwidth}{0.5em}
  \setlength{\labelsep}{0em} } }

\newcommand{\squishend}{
 \end{list} }

\AtBeginDocument{%
  \providecommand\BibTeX{{%
    \normalfont B\kern-0.5em{\scshape i\kern-0.25em b}\kern-0.8em\TeX}}}

\usepackage{etoolbox,refcount}

\newcounter{countitems}
\newcounter{nextitemizecount}
\newcommand{\setupcountitems}{%
  \stepcounter{nextitemizecount}%
  \setcounter{countitems}{0}%
  \preto\item{\stepcounter{countitems}}%
}
\makeatletter
\newcommand{\computecountitems}{%
  \edef\@currentlabel{\number\c@countitems}%
  \label{countitems@\number\numexpr\value{nextitemizecount}-1\relax}%
}
\newcommand{\nextitemizecount}{%
  \getrefnumber{countitems@\number\c@nextitemizecount}%
}
\newcommand{\previtemizecount}{%
  \getrefnumber{countitems@\number\numexpr\value{nextitemizecount}-1\relax}
}
\makeatother
{\end{itemize}%
\unskip\computecountitems\ifnumcomp{\previtemizecount}{>}{3}{\end{multicols}}{}}

\newcommand\opstyle{\color{red}}
\lstdefinestyle{customcpp}{
  belowcaptionskip=1\baselineskip,
  xleftmargin=17pt,
  xrightmargin=3pt,
  language=C++,
  showstringspaces=false,
  basicstyle=\scriptsize\ttfamily,
  keywordstyle=\bfseries\color{purple!40!black},
  commentstyle=\itshape\color{blue},
  identifierstyle=\color{black},
  stringstyle=\color{orange},
  morekeywords={~!+-*&^/\%},
  numbers=left,
  stepnumber=1,
}

\begin{document}

\title{Deep Learning based Vulnerability Detection: \\ Are We There Yet?}

\author{Saikat Chakraborty, Rahul Krishna, Yangruibo Ding, Baishakhi Ray
\IEEEcompsocitemizethanks{
\IEEEcompsocthanksitem Chakraborty, S., Krishna, R., Ding, Y., and Ray, B., are with Columbia University, New York, NY, USA.\protect\\
E-mail: saikatc@cs.columbia.edu, i.m.ralk@gmail.com, yangruibo.ding@columbia.edu, and rayb@cs.columbia.edu.}
}

\markboth{IEEE TRANSACTIONS ON SOFTWARE ENGINEERING, VOL. TBD, 2020}{Chakraborty \etal: Deep Learning Based Vulnerability Detection: Are We There Yet?}

\IEEEtitleabstractindextext{%
\begin{abstract}

Automated detection of software vulnerabilities is a fundamental problem in software security. Existing program analysis techniques either suffer from high false positives or false negatives. Recent progress in Deep Learning (DL) has resulted in a surge of interest in applying DL for automated vulnerability detection. Several recent studies have demonstrated promising results achieving an accuracy of up to 95\% at detecting vulnerabilities. 
In this paper, we ask, \textit{``how well do the state-of-the-art DL-based techniques perform in a real-world \vp scenario?''}. To our surprise, we find that their performance drops by more than 50\%.
A systematic investigation of what causes such precipitous performance drop reveals that existing DL-based \vp approaches suffer from challenges with the training data (\eg data duplication, unrealistic distribution of vulnerable classes, etc.) and with the model choices (\eg simple token-based models). As a result, these approaches often do not learn features related to the actual cause of the vulnerabilities. Instead, they learn unrelated artifacts from the dataset (\eg specific variable/function names, etc.). 
Leveraging these empirical findings, we demonstrate how a more principled approach to data collection and model design, based on realistic settings of \vp, can lead to better solutions. The resulting tools perform significantly better than the studied baseline\textemdash up to 33.57\% boost in precision and 128.38\% boost in recall compared to the best performing model in the literature. Overall, this paper elucidates existing DL-based \vp systems' potential issues and draws a roadmap for future DL-based \vp research. In that spirit, we make available all the artifacts supporting our results:  \url{https://git.io/Jf6IA}.
  
\end{abstract}

}


\maketitle

\IEEEdisplaynontitleabstractindextext

\IEEEpeerreviewmaketitle


\section{Introduction}
\label{sec:intro}

Automated detection of security vulnerabilities is a fundamental problem in systems security.  
Traditional techniques are known to suffer from high false-positive/false-negative rates~\cite{johnson2013don, smith2015questions,ayewah2007evaluating, newsome2005dynamic, liu2012software}.
For example, static analysis-based tools typically result in high false positives, \ie, detect non-vulnerable cases as vulnerable, and dynamic analysis suffers from high false negatives, \ie, cannot detect many real vulnerabilities. After prolonged effort, these tools remain unreliable, leaving significant manual overhead for developers~\cite{smith2015questions}.

Recent progress in Deep Learning (DL), especially in domains like computer vision and natural language processing, has sparked interest in using DL to detect security vulnerabilities automatically with high accuracy. According to Google scholar, 92 papers appeared in popular security and software engineering venues  between  2019  and  2020 that apply learning techniques to detect different types of bugs\footnote{published in TSE, ICSE, FSE, ASE, S\&P Oakland, CCS, USENIX Security, etc.}. In fact, several recent studies have demonstrated very promising results achieving high accuracy (up to $95\%$) at detecting vulnerabilities~\cite{li2018vuldeepecker, li2018sysevr, russell2018automated, li2017large, maiorca2019digital, suarez2017droidsieve, zhou2019devign}. 

Given such remarkable reported success of DL models at detecting vulnerabilities, it is natural to ask why they are performing so well, what kind of features these models are learning, and most importantly, whether they can be used effectively and reliably in detecting real-world vulnerabilities.  Understanding such explainability and generalizability of the DL models is pertinent as it may help solve similar problems in other domains like computer vision~\cite{barreno2006can, torralba2011unbiased}.

For instance, the generalizability of a DL model is limited by implicit biases in the dataset, which are often introduced during the dataset generation/curation/labeling process and therefore affect both the testing and training data equally (assuming that they are drawn from the same dataset). These biases tend to allow DL models to achieve high accuracy in the test data by learning highly idiosyncratic features specific to that dataset instead of generalizable features. For example, Yudkowsky et al.~\cite{yudkowsky2008artificial} described an instance where US Army found out that a neural network for detecting camouflaged tanks did not generalize well due to dataset bias even though the model achieved very high accuracy in the testing data. They found that all the photos with the camouflaged tanks in the dataset were shot in cloudy days, and the model simply learned to classify lighter and darker images instead of detecting tanks.

In this paper, we systematically measure the generalizability of four state-of-the-art Deep Learning-based Vulnerability Prediction (hereafter \dlvp) techniques~\cite{li2018vuldeepecker, li2018sysevr, russell2018automated, zhou2019devign} that have been reported to detect security vulnerabilities with high accuracy (up to $95\%$) in the existing literature. We primarily focus on the Deep Neural Network (DNN) models that take source code as input~\cite{li2018vuldeepecker, li2018sysevr, russell2018automated, zhou2019devign, yamaguchi2014modeling} and detect vulnerabilities at function granularity. These models operate on a wide range of datasets that are either generated synthetically or adapted from real-world code to fit in simplified \vp settings. 

First, we curate a new vulnerability dataset 
from two large-scale popular real-world projects (Chromium and Debian) to evaluate the performance of existing techniques in the real-world \vp setting. The code samples are annotated as vulnerable/non-vulnerable, leveraging their issue tracking systems. Since both the code and annotations come from the real-world, detecting vulnerabilities using such a dataset reflects a realistic vulnerability prediction scenario. We also use \devigndata dataset proposed by Zhou~\etal~\cite{zhou2019devign}.

To our surprise, we find that none of the existing models perform well in real-world settings. If we directly use a pre-trained model to detect the real-world vulnerabilities, the performance drops by $\sim$\textit{73\%}, on average. Even if we retrain these models with real-world data, their performance drops by $\sim$\textit{54\%} from the reported results.
For example, VulDeePecker~\cite{li2018vuldeepecker} reported a precision of 86.9\% in their paper. However, when we use VulDeePecker's pre-trained model in real world datasets, its precision reduced to 11.12\%, and after retraining, the precision becomes 17.68\%. A thorough investigation of such poor performance reveals several problems:
\bi
\item 
\textit{Inadequate Model.} The most popular models are token-based, which treat code as a sequence of tokens and do not take into account semantic dependencies that play a vital role in vulnerability predictions. Even when a graph-based model is used, it does not focus on increasing the class-separation between vulnerable and non-vulnerable categories. Thus, in realistic scenarios, they suffer from low precision and recall. 

\item
\textit{Learning Irrelevant Features.} 
While looking at the features that the existing techniques are picking up (using state-of-the-art explanation techniques~\cite{springenberg2014striving,guo2018lemna}), we find that the state-of-the-art models are essentially picking up irrelevant features that are not related to vulnerabilities and are likely artifacts of the training datasets. 

\item \textit{Data Duplication.} The training and testing data in most existing approaches contain duplicates (up to \textit{68\%}); thus, artificially inflating the reported results.

\item \textit{Data Imbalance.} Existing approaches do not alleviate the class imbalance problem~\cite{sun2009classification,seiffert2009rusboost} of real-world vulnerability distribution as non-vulnerable code is much more frequent than the vulnerable ones.

\ei
Having established these concerns empirically, we propose a road-map that we hope will help the DL-based vulnerability prediction researchers to avoid such pitfalls in the future.  
To this end, we demonstrate how a more principled approach to data collection and model design, based on our empirical findings, can lead to better solutions. 
For data collection, we discuss how to curate real-world \vp data incorporating both static and evolutionary (\ie, bug-fix) nature of the vulnerabilities.
For model building, we show representation learning~\cite{6472238} can be used on top of traditional DL methods to increase the class separation between vulnerable and non-vulnerable samples. 
Representation learning is a popular class of machine learning techniques that automatically discovers the input representations needed for improving classification, and thus, replaces the need for manual feature engineering. Our key insight is as follows: distinguishing features of vulnerable and benign code is complex; thus, the model must learn to represent  them automatically in the feature space.

We further empirically establish that using semantic information (with graph-based models), data de-duplication, and balancing training data to address the class imbalance of vulnerable/non-vulnerable samples can significantly improve \vp. 
Following these steps, we can boost precision and recall of the best performing model in the literature by up to 33.57\% and 128.38\%  respectively over current baselines.

In summary, our contributions in this paper are:

\begin{enumerate}
\item We systematically study existing approaches in \dlvp task and identify several problems with the current dataset and modeling practices. 


\item Leveraging the empirical results, we propose a summary of best practices that can help future \dlvp research and experimentally validate these suggestions.  

\item We curated a real-world dataset from developer/user reported vulnerabilities of Chromium and Debian projects. We release our dataset in this anonymous directory \url{https://bit.ly/3bX30ai}.

\item We also open source all our code and data we used in this study for broader dissemination. Our code and replication data are available in \url{https://git.io/Jf6IA}.
\end{enumerate}
To this end, we argue that DL-based vulnerability detection is still very much an open problem and requires a well-thought-out data collection and model design framework guided by real-word vulnerability detection settings.
\section{Background and Challenges}
\label{sect:background}

\dlvp methods aim to detect unknown vulnerabilities in target software by learning different vulnerability patterns from a training dataset. Most popular \dlvp approaches consist of three steps: data collection, model building, and evaluation. 
First, data is collected for training, and an appropriate model is chosen as per design goal and resource constraints.
The training data is preprocessed according to the format preferred by the chosen model. 
Then the model is trained to minimize a loss function. The trained model is intended to be used in the real world. 
To assess the effectiveness of the model performance of the model is evaluated on unseen test examples.

This section describes the theory of DL-based \vp approaches (\S\ref{subsec:overview_models}), existing datasets (\S\ref{subsec:dataset_and_limitation}), existing modeling techniques (\S\ref{subsec:model_and_limitation}), and evaluation procedure (\S\ref{subsec:evaluation_and_limilation}). Therein, we discuss the challenges that potentially limit the applicability of existing \dlvp techniques. 

\subsection{\dlvp Theory}
\label{subsec:overview_models}


DL-based vulnerability predictors learn the vulnerable code patterns from a training data ($D_{train}$) set where code elements are labeled as vulnerable or non-vulnerable.  Given a code element $(x)$ and corresponding vulnerable/non-vulnerable label $(y)$, the goal of the model is to learn features that maximize the probability $p(y|x)$ with respect to the model parameters ($\theta$). Formally, training a model is learning the optimal parameter settings ($\theta^*$) such that,
\begin{equation}
    \theta^* = argmax_\theta{\prod\limits_{(x,y) \in D_{train}}p(y|x, \theta})
    \label{eqn:prob}
\end{equation}
First, a code element ($x^i$) is transformed to a real valued vector ($h^i \in \mathbb{R}^n$), which is a compact representation of $x^i$. How a model transforms $x^i$ to $h^i$ depends on the specifics of the model. This $h^i$ is transformed to a scalar $\hat{y} \in [0, 1]$ which denotes the probability of code element $x^i$ being vulnerable. In general, this transformation and probability calculation is achieved through a feed forward layer and a softmax~\cite{bridle1990probabilistic} layer in the model. Typically, for binary classification task like \vp, optimal model parameters are learned by minimizing the cross-entropy loss~\cite{suter1990multilayer}. Cross-entropy loss penalizes the discrepancy in the model's predicted probability and the actual probability (0. for non-vulnerable 1. for vulnerable examples)~\cite{plunkett1997exercises}. 






\begin{figure}[!htb]
    \centering
    \includegraphics[width=.65\linewidth]{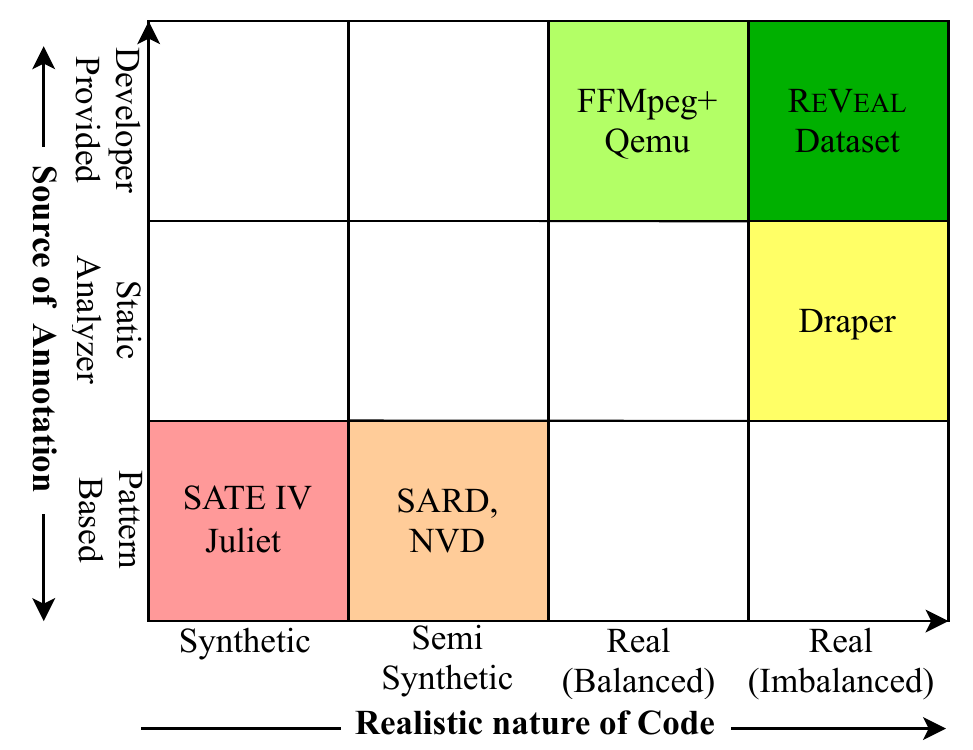}  
    \caption{{\small{ Different \dlvp dataset and their synthetic/realistic nature.} From \lone{red} to \lsix{green}, colors symbolize increasing realistic nature of dataset. \lone{Red} is the most synthetic, \lsix{green} is the most realistic.
    }}
    \label{fig:realism_axis}
\end{figure}

\subsection{Existing Dataset}
\label{subsec:dataset_and_limitation}
To train a \vp model, we need a set of annotated code with labels vulnerable or benign. The number of vulnerable code should be large enough to allow the model to learn from it. Researchers used a wide spectrum of data sources to collect data for \dlvp (see~\Cref{fig:realism_axis}). 
Depending on how the code samples are collected and how they are annotated, we classify them as: 
\bi
\item \textit{Synthetic data:} The vulnerable code example and the annotations are artificially created. SATE IV Juliet~\cite{okun2013report} dataset and SARD~\cite{SARD} fall in this category. Here the examples are synthesized using known vulnerable patterns. These datasets were originally designed for evaluating traditional static and dynamic analysis based \vp tools. 

\item \textit{Semi-synthetic data:} Here either the code or the annotation is derived artificially. 
For example, Draper dataset, proposed by Russell~\etal~\cite{russell2018automated}, contains functions that are collected from open source repositories but are annotated using static analyzers. 
Examples of SARD~\cite{SARD} and National Vulnerability Database (NVD ~\cite{booth2013national}) dataset are also taken from production code; however, they are often modified in a way to demonstrate the vulnerability isolating them from their original context. Although these datasets are more complex than synthetic ones, they do not fully capture the complexities of the real-world vulnerabilities due to simplifications and isolations. 

\item \textit{Real data:} Here both the code and the corresponding vulnerability annotations are derived from real-world sources. For instance, Zhou~\etal~\cite{zhou2019devign} curated \textit{Devign} dataset, which consists of past vulnerabilities and their fixes from four open-source projects, two of which are publicly available. 

\ei

\noindent
\textbf{Limitations.} The problems with the dataset lie in how realistic the data source is and  how they are annotated (see~\Cref{fig:realism_axis}). A model trained on a synthetic dataset comprising of simple patterns will be limited to detecting only those simple pattern which seldom occur in real life. 
For instance, consider an atypical {buffer overflow} example  in~\Cref{code:simple_synthetic_code} used by VulDeePecker and SySeVR. Albeit a good pedagogical example, real world vulnerabilities are not as simple or as isolated. Figure~\ref{code:real_linux_code} shows another buffer overflow example from linux kernel. Though the fix is very simple, finding the vulnerability itself requires an in-depth reasoning about the semantics of different components (\ie, variables, functions etc.) of the code. A model is trained to reason about simpler examples as in Figure~\ref{code:simple_synthetic_code} will fail to reason about Figure~\ref{code:real_linux_code} code. Further, any model annotated by a static analyzer~\cite{russell2018automated} inherits all the drawbacks, \eg, high false positive rate~\cite{johnson2013don, smith2015questions}.
\begin{figure}[!tp]
\begin{lstlisting}
void action(char *data) const {
// FLAW: Increment of pointer in the loop will cause
// freeing of memory not at the start of the buffer.
    for (; *data != '\0'; data++){
        if (*data == SEARCH_CHAR){
            printLine("We have a match!");
            break;
        }
    }
    free(data);
}
\end{lstlisting}
\caption{\small {Example Vulnerability (CWE761)}~\cite{cwe761_vd}.}
\label{code:simple_synthetic_code}
\end{figure}
\begin{figure}[!tp]
\scriptsize
\begin{lstlisting}
static void eap_request(
        eap_state *esp, u_char *inp, int id, int len) {
	...
	if (vallen < 8 || vallen > len) {
			...
			break;
	}
/* FLAW: 'rhostname' array is vulnerable to overflow.*/
@\red{\textbf{-}}@	@\red{\textbf{if (vallen >= len + sizeof (rhostname))\{}}@
@\green{\textbf{+}}@	@\green{\textbf{if (len - vallen >= (int)sizeof (rhostname))\{}}@
        ppp_dbglog(...);
        MEMCPY(rhostname, inp + vallen, sizeof(rhostname) - 1);
        rhostname[sizeof(rhostname) - 1] = '\0';
        ...
    }
    ...
}
\end{lstlisting}
\caption{\small {CVE-2020-8597 - A partial patch (original patch~\cite{linux_pppd_vuln_patch}) for an instance of buffer overflow vulnerability in linux point to point protocol daemon ({\tt pppd}) due to a logic flaw in the packet processor}~\cite{linux_pppd_vuln_analysis_1, linux_pppd_vuln_analysis_2}. 
}
\label{code:real_linux_code}
\end{figure}
In the most realistic dataset, \devigndata~\cite{zhou2019devign}, the ratio of vulnerable and non-vulnerable examples is approximately 45\%-55\%, which does not reflect the real world distribution of vulnerable code. Further, the dataset only contains function that annotates functions that went through vulnerability-fix commits as vulnerable. When a model is trained on such dataset, the model is not presented with other functions from a vulnerable functions' context, thus will not be as effective in differentiating vulnerable functions from other non-vulnerable functions from the context.


\subsection{Existing Modeling Approaches}
\label{subsec:model_and_limitation}
Model selection depends primarily on the information that one wants to incorporate. The popular choices for \dlvp are token-based or graph-based models, and the input data (code) is preprocessed accordingly~\cite{li2018vuldeepecker, russell2018automated, zhou2019devign}. 

\bi
\item \textit{Token-based models:}
In the token-based models, code is considered as a sequence of tokens.
Existing token-based models used different Neural Network architectures. For instance, Li~\etal~\cite{li2018vuldeepecker} proposed a Bidirectional Long Short Term Memory (BSLTM) based model, Russell~\etal~\cite{russell2018automated} proposed a Convolutional Neural Network (CNN) and Radom Forest-based model and compared against Recurrent Neural Network (RNN) and CNN based baseline models for \vp. For these relatively simple token-based models, token sequence length is an important factor to impact performance as it is difficult for the models to reason about long sequences. To address this problem, VulDeePecker~\cite{li2018vuldeepecker} and SySeVR~\cite{li2018sysevr} extract code slices. The motivation behind slicing is that not every line in the code is equally important for \vp. Therefore, instead of considering the whole code, only slices extracted from ``interesting points'' in code (\eg API calls, array indexing, pointer usage, etc.) are considered for vulnerability prediction and rest are omitted. 

\item \textit{Graph-based models:}
These models consider code as graphs and incorporate different syntactic and semantic dependencies. Different type of syntactic graph (Abstract Syntax Tree) and semantic graph (Control Flow graph, Data Flow graph, Program Dependency graph, Def-Use chain graph etc.) can be used for \vp. 
For example, Devign~\cite{zhou2019devign} leverage code property graph (CPG) proposed by Yamaguchi~\etal~\cite{yamaguchi2014modeling} to build their graph based \vp model. CPG is constructed by augmenting different dependency edges (\ie, control flow, data flow, def-use, etc.) to the code's Abstract Syntax Tree (AST)  (see~\tion{pipeline} for details).
\ei
Both graph and token-based models have to deal with {\em vocabulary explosion} problem\textemdash  the number of possible identifiers (variable, function name, constants) in code can be virtually infinite, and the models have to reason about such identifiers.  A common way to address this issue is to replace the tokens with abstract names~\cite{li2018vuldeepecker, li2018sysevr}. For instance, VulDeePecker~\cite{li2018vuldeepecker} replaces most of the variable and function names with symbolic names ({\tt VAR1, VAR2, FUNC1, FUNC2} etc.).

Expected input for all the models are real valued vectors commonly known as embeddings. There are several ways to embed tokens to vectors. One such way is to use an embedding layer~\cite{turian2010word} that is jointly trained with the \vp task~\cite{russell2018automated}. Another option is to use external word embedding tool(\eg \emph{Word2Vec}~\cite{rong2014word2vec}) to create vector representation of every token. VulDeePecker~\cite{li2018vuldeepecker} and SySeVR~\cite{li2018sysevr} uses \emph{Word2Vec} to transform their symbolic tokens into vectors. Devign~\cite{zhou2019devign}, in contrast, uses \emph{Word2Vec} to transform the concrete code tokens to real vectors. 

Once a model is chosen and appropriate preprocessing is done on the training dataset, the model is ready to be trained by minimizing a loss function. Most of the existing approaches optimize the model by minimizing some variation of cross-entropy loss. For instance, Russell~\etal~\cite{russell2018automated} optimized their model using cross-entropy loss, Zhou~\etal~\cite{zhou2019devign} used regularized cross entropy loss. 

\begin{figure}[!htpb]
\centering
\includegraphics[width=0.80\linewidth]{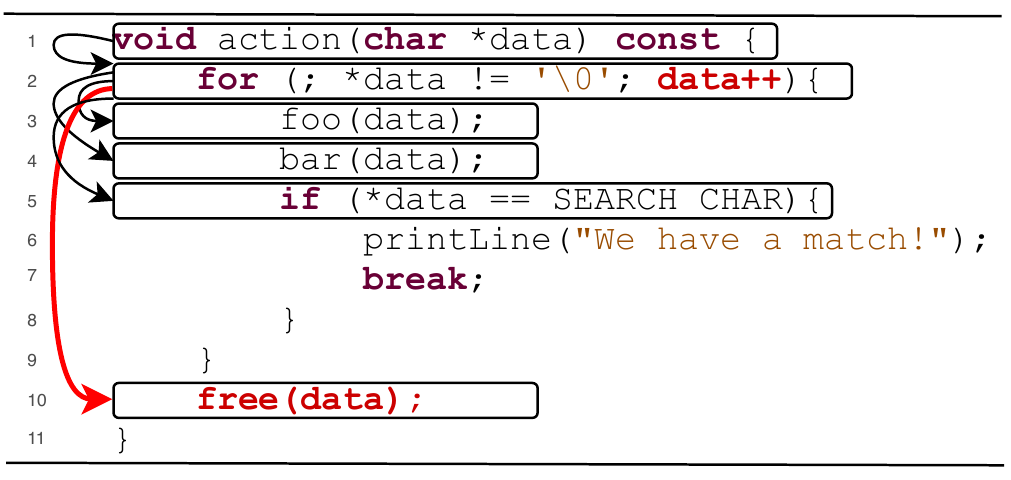}
\caption{\small {Example of CWE-761~\cite{cwe761}}. A buffer is freed not at the start of the buffer but somewhere in the middle of the buffer. This can cause the application to crash, or in some cases, modify critical program variables or execute code. This vulnerability can be detected with data dependency.
}
\label{code:token_vuln}
\end{figure}

\noindent
\textbf{Limitations.} 
Token based models assume that tokens are linearly dependent on each other, and thus, only lexical dependencies between the tokens are present, while the semantic dependencies are lost, which often play important roles in \vp~\cite{clang, cppcheck, flawfiner}. 
To incorporate some semantic information, VulDeePecker~\cite{li2018vuldeepecker} and SySeVR~\cite{li2018sysevr} extracted program slices of a potentially interesting point. For example, consider the code in Figure~\ref{code:token_vuln}. A slice \wrt {\tt free} function call at line 10 gives us all the lines except lines 6 and 7. The token sequence of the slice are: {\tt void action ( char * data ) const \{ for ( data ; * data != `$\backslash$0' ; \textbf{data ++} ) \{ foo ( data ) ; bar ( data ) ; if ( * data == SEARCH\_CHAR ) \{ \textbf{free ( data )} ;}. In this examples, while the two main components for this code being vulnerable, \ie~{\tt \textbf{data ++}} (line 2) ~and {\tt \textbf{free ( data )}} (line 10) are present in the token sequence, 
they are far apart from each other without explicitly maintaining any  dependencies. 

In contrast, as a graph based model can consider 
the data dependency edges (\red{red} edge), we see that there is a direct edge between those lines making those lines closer to each other making it easier for the model to reason about that connection. Note that this is a simple CWE example (CWE 761), which requires only the data dependency graph to reason about. Real-world vulnerabilities are much more complex and require reasoning about control flow, data flow, dominance relationship, and other kinds of dependencies between code elements~\cite{yamaguchi2014modeling}. However,  graph-based models, in general, are much more expensive than their token-based counterparts and do not perform well in a resource-constrained environment. 

One problem with the existing approaches is that although the trained models learn to discriminate vulnerable and non-vulnerable code samples, the training paradigm does not explicitly focus on increasing the  separation between the vulnerable and non-vulnerable examples.
Thus, with slight variations the classifications become brittle. 

Another problem pertains to data imbalance~\cite{wu2003class} between vulnerable and benign code as the proportion of vulnerable examples in comparison to the non-vulnerable one in real world dataset is extremely low~\cite{russell2018automated}.
When a model is trained  on such imbalanced dataset, models tend to be biased by the non-vulnerable examples. 

\subsection{Existing Evaluation Approaches}
\label{subsec:evaluation_and_limilation}
To understand the applicability of a trained model for detecting vulnerability in the real-world, it must first be evaluated. In most cases, a trained model is evaluated on held out test set. Test examples go through the same pre-processing technique as the training and then the model predicts the vulnerability of those pre-processed test examples. This evaluation approach gives an estimate of how the model may perform when used to detect vulnerabilities in the real-world. 

\noindent
\textbf{Limitations.} Although all the existing approaches report their performances using their own evaluation dataset, it does not give a comprehensive overview of the applicability of the model in the real-world. All we can learn from such intra-dataset evaluation is how well their approach fits their own dataset. 
Although there are some limited case studies on such models finding vulnerabilities in real-world projects, those case studies do not shed light on the false positives and false negatives~\cite{johnson2013don}. 
The number of false positives and false negatives are directly correlated to the developer effort in \vp~\cite{aggarwal2006integrating} and too much of any would hold the developer from using the model~\cite{heckman2011systematic, muske2016survey}.

\begin{table*}[!tpb]
\centering
\caption{{\small {Summary of \dlvp datasets and approaches.}}}
\resizebox{0.95\textwidth}{!}{
\begin{tabular}{l|l|r|r|l|l|l|l}
\hlineB{2}
\textbf{Dataset}  & \textbf{Used By} & \textbf{\# Programs} & \textbf{\% Vul*} & \textbf{Granularity} & \textbf{Model Type} & \textbf{Model} & \textbf{Description}\bigstrut\\
\hlineB{2}
SATE IV Juliet~\cite{okun2013report}  & Russell~\etal~\cite{russell2018automated} & 11,896 & 45.00 & Function & Token & CNN+RF & Synthetic code for testing static analyzers.\bigstrut\\
\hline
\multirow{2}{*}{SARD~\cite{SARD}}  & VulDeePecker~\cite{li2018vuldeepecker} & 9,851 & 31 & Slice & Token & BLSTM & \multirow{2}{*}{\begin{tabular}[c]{@{}l@{}}Synthetic, academic, and production security \bigstrut\\ flaws or vulnerabilities.\end{tabular}} \bigstrut\\
\cline{2-7}
& SySeVR~\cite{li2018sysevr} & 14,000 & 13.41 & Slice & Token & BGRU & \bigstrut\\
\hline
\multirow{2}{*}{NVD~\cite{booth2013national}} & VulDeePecker$^\ddagger$ & 840 & 31 & Slice & Token & BLSTM & \multirow{2}{*}{\begin{tabular}[c]{@{}l@{}}Collection of known vulnerabilities \bigstrut\\ from real world projects.\end{tabular}} \bigstrut\\
\cline{2-7}
& SySeVR$^\ddagger$ & 1,592 & 13.41 & Slice & Token & BGRU & \bigstrut\\
\hline
Draper~\cite{russell2018automated} & Russell~\etal~\cite{russell2018automated} & 1,274,366  & 6.46  & Function  & Token & CNN+RF & \begin{tabular}[c]{@{}l@{}}Contains   code   from   public   repositories \bigstrut\\ in  Github  and  Debian  source repositories.\bigstrut\\\end{tabular} \bigstrut\\
\hline
\devigndata~\cite{zhou2019devign} & Devign~\cite{zhou2019devign} & 22,361 & 45.02 & Function  & Graph & GGNN & \begin{tabular}[c]{@{}l@{}}FFMPeg   is   a   multimedia   library;\bigstrut\\ Qemu is hardware virtualization emulator.\bigstrut\\\end{tabular} \bigstrut\\
\hline
\tool dataset   & This paper  & 18,169 & 9.16  & Function  & Graph & \begin{tabular}[c]{@{}l@{}}GGNN + \bigstrut\\ MLP + \bigstrut\\ Triplet Loss\end{tabular} & \begin{tabular}[c]{@{}l@{}}Contains code from Chromium and Debian \bigstrut\\ source code repository\\\end{tabular}\bigstrut\\
\hlineB{2}
\multicolumn{8}{l}{* ~ Percentage of vulnerable samples in the dataset.}\bigstrut\\
\multicolumn{8}{l}{$^\ddagger$ ~ VulDeePecker and SySeVR   uses combination of SARD and NVD datasets to train and evaluate their model.}
\end{tabular}
}
\label{tab:existing_approach_design_choices}
\end{table*}

\section{\tool Data Collection}
\label{sect:tool_data_collection_section3}

\begin{figure}[t!]
    \centering
    \includegraphics[width=0.75\linewidth]{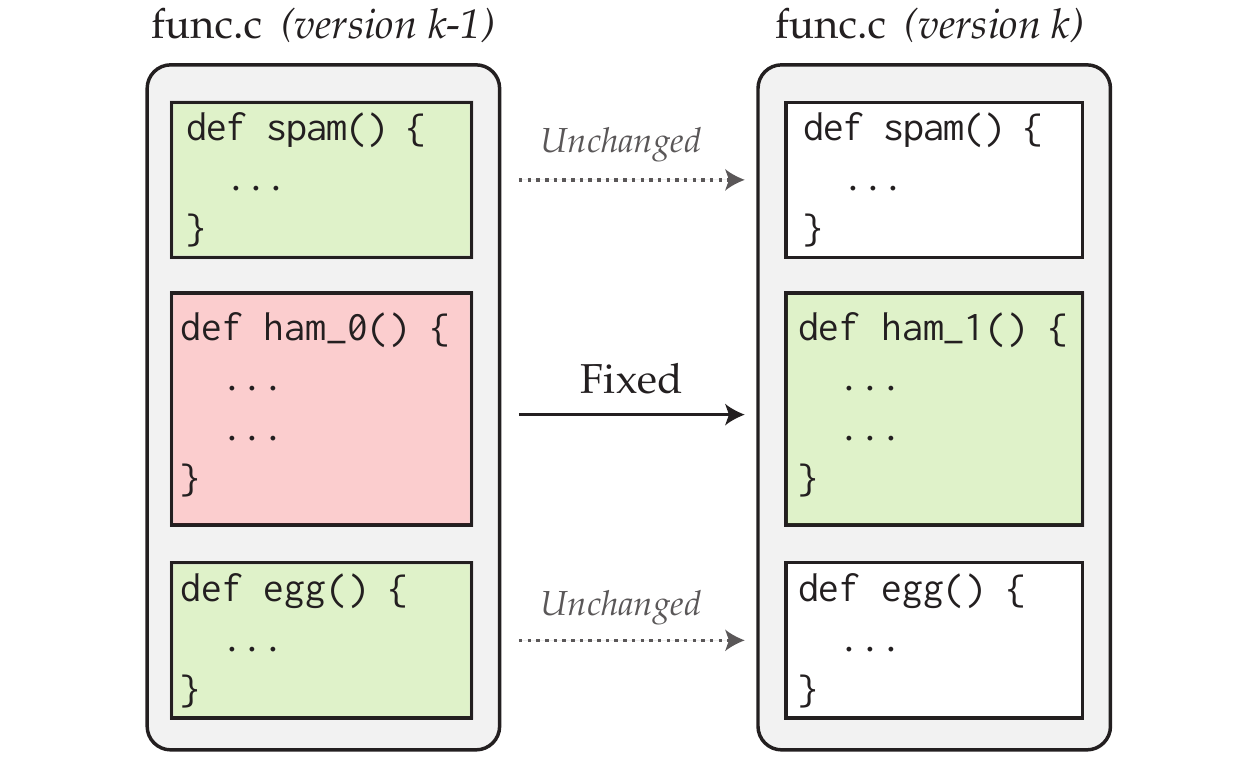}
    \caption{\small {Collecting real world data for \tool.} Green samples are labeled as \emph{non-vulnerable}, while red sample is marked as \emph{vulnerable}.}
    \label{fig:data_collection}
\end{figure}


To address the limitations with the existing data sets, we curate a more robust and comprehensive real world dataset, \tool, by tracking the past vulnerabilities from two open-source projects: Linux Debian Kernel and Chromium (open source project of Chrome). We select these projects because:
(i) these are two popular and well-maintained public projects with large evolutionary history, 
(ii) the two projects represent two important program domains (OS and browsers) that exhibit diverse security issues, and
(iii) both the projects have plenty of publicly available vulnerability reports.

To curate our data, we first collect \emph{already fixed} issues with publicly available patches. For Chromium, we scraped its bug repository Bugzilla\footnote{\url{https://bugs.chromium.org/p/chromium/issues/list}}. For Linux Debian Kernel, we collected the issues from Debian security tracker\footnote{\url{https://security-tracker.debian.org/tracker/}}.  We then identify vulnerability related issues, \ie, we choose those patches that are labeled with ``security''. This identification mechanism is inspired by the security issue identification techniques proposed by Zhou~\etal~\cite{zhou2017automated}, where they filter out commits that do not have security related keywords.

For each patch, we extracted the corresponding vulnerable and fixed versions (\ie, old and new version) of {\tt C/C++} source and header files that are changed in the patch. We annotate the previous versions of all changed functions (\ie, the versions prior to the patch) as vulnerable and the fixed version of all the changed functions (\ie, the version after patch) as `clean'.  Additionally, other functions that were not involved in the patch (\ie, those that remained unchanged) are all annotated as `clean'. 

A contrived example of our data collection strategy is illustrated in~\fig{data_collection}. Here, we have two versions of a file \texttt{file.c}. The previous version of the file (version $k-1$) has a vulnerability which is fixed in the subsequent version (version $k$) by patching the function \texttt{ham\_0()} to \texttt{ham\_1()}. In our dataset, \texttt{ham\_0()} would be included and labeled `vulnerable' and \texttt{ham\_1()} would be included and labeled `clean'. The other two functions (\texttt{spam()} and \texttt{egg}) remained unchanged in the patch. Our dataset would include a copy of these two functions and label them as `clean'.

Annotating code in this way simulates real-world \vp scenario, where a DL model would learn to inspect the vulnerable function in the context of all the other functions in its scope. Further, by retaining the fixed variant of the vulnerable function, the DL model may learn the nature of patch. 
We make available our data collection framework
and the curated vulnerability data for \textit{Chromium} and \textit{Debian}\footnote{Chromium and Debian dataset:~\url{https://bit.ly/3bX30ai}} for broader dissemination.

\begin{figure*}[t!]
    \centering
    \includegraphics[width=\linewidth, keepaspectratio = true]{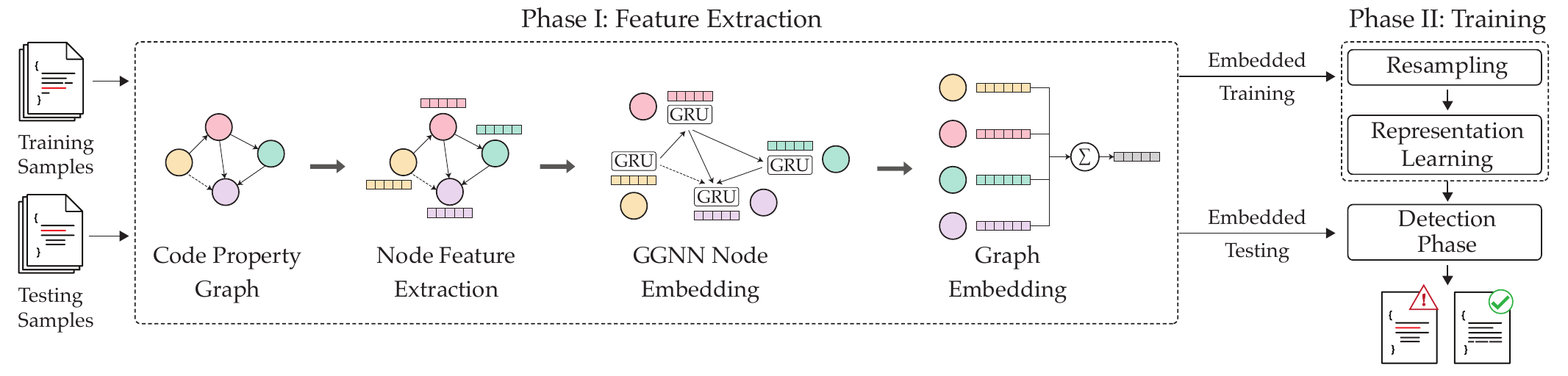}
    \caption{\small {Overview of the \tool vulnerability prediction framework.}}
    \label{fig:intro_fig}
\end{figure*}

\section{\tool Pipeline}
\label{sect:pipeline}

\begin{figure}[tp!]
    \algoframework\vspace{0.66em}
    \algofeature
\end{figure}




In this section, we present a brief overview of the \tool pipeline that aims to more accurately detect the presence of real-world vulnerabilities.~\fig{intro_fig} illustrates the \tool pipeline. It operates in two phases namely, feature extraction (Phase-I) and training (Phase-II). In the first phase we translate real-world code into a graph-embedding (\tion{feature_extraction}). In the second phase, we train a representation learner on the extracted features to learn a representation that most ideally demarcates the vulnerable examples from non-vulnerable examples (\tion{training}). Algorithm~\ref{alg:overview_psuedocode} shows the full training procedure for \tool. This algorithm expects Training data -- a list of tuples, where each tuple contains a code ($C$) and corresponding vulnerability annotation ($l$). 




\subsection{Feature Extraction (Phase-I)}
\label{sect:feature_extraction}

The goal of this phase is to convert code into a compact and a uniform length feature vector while maintaining the semantic and syntactic information. 
Note that, the feature extraction scheme presented below represents the most commonly used series of steps for extracting features from a graph representation~\cite{zhou2019devign}. \tool uses Algorithm~\ref{alg:feature_extraction} to extract graph embedding (graph based feature vector that represent the entirety of a function in a code).

To extract the syntax and semantics in the code, we generate a code property graph (hereafter, CPG)~\cite{yamaguchi2014modeling}. The CPG is a particularly useful representation of the original code since it offers a combined and a succinct representation of the code consisting of elements from the control-flow and data-flow graph in addition to the AST and program dependency graph (or PDG). Each of the above elements offer additional context about the overall semantic structure of the code~\cite{yamaguchi2014modeling}.

Formally, a CPG is denoted as $\mathrm{G}=\left(\mathrm{V}, \mathrm{E}\right)$, where $\mathrm{V}$ represent the vertices (or nodes) in the graph and $\mathrm{E}$ represents the edges. Each vertex $\mathrm{V}$ in the CPG is comprised of the vertex type (\eg~{\tt ArithmeticExpression, CallStatement} etc.) and a fragment of the original code. To encode the type information, we use a one-hot encoding vector denoted by $T_v$. To encode the code fragment in the vertex, we use a word2vec embedding denoted by $C_v$. Next, to create the vertex embedding, we concatenate $T_v$ and $C_v$ into a joint vector notation for each vertex.

The current vertex embedding is not adequate since it considers each vertex in isolation. It therefore lacks information about its adjacent vertices and, as a result, the overall graph structure. This may be addressed by ensuring that each vertex embedding reflects both its information and those of its neighbors. We use gated graph neural networks (hereafter GGNN)~\cite{li2015gated} for this purpose.

Feature vectors for all the nodes in the graph ($X$) along with the edges ($\mathrm{E}$) are the input to the GGNN~\cite{li2015gated, zhou2017automated}. For every vertex in the CPG, GGNN assigns a gated recurring unit (GRU) that updates the current vertex embedding by assimilating the embedding of all its neighbors. Formally, 

$$x_v' = GRU(x_v, \sum\limits_{(u, v) \in \mathrm{E}}{g(x_u)})$$

Where, $GRU(\cdot)$ is a Gated Recurrent Function, $x_v$ is the embedding of the current vertex $v$, and $g(\cdot)$ is a transformation function that assimilates the embeddings of all of vertex $v$'s neighbors~\cite{zhao2017machine, li2015gated, allamanis2017learning}. $x_v^\prime$ is the GGNN-transformed representation of the vertex $v$'s original embedding $x_v$. $x_v^\prime$ now incorporates $v$'s original embedding $x_v$ as well as the embedding of its neighbors. 

The final step in preprocessing is to aggregate all the vertex embedding $x_v^\prime$ to create a single vector representing the whole CPG denoted by $x_g$, \ie:

\begin{equation*}
    x_g = \sum\limits_{v\in\mathrm{V}}{x_v^\prime}
\end{equation*}

Note that \tool uses a simple element-wise summation as the aggregation function, but in practice it is a configurable parameter in the pipeline. The result of the pipeline presented so far is an $m-$dimensional feature vector representation of the original source code. To pre-train the GGNN, we augment a classification layer on top of the GGNN feature extraction. This training mechanism is similar to Devign~\cite{zhou2019devign}. Such pre-training deconstructs the task of ``learning code representation'', and ``learning vulnerability'', and is also used by Russell~\etal~\cite{russell2018automated}. While, we pre-train GGNN in a supervised fashion, unsupervised program representation learning~\cite{lachaux2020unsupervised} can also be done to learn better program presentation. However, such learning is beyond the scope of this research and we leave that for future research.  


\subsection{Training (Phase-II)}
\label{sect:training}

In real-world data, the number of non-vulnerable samples (\ie, negative examples) far outnumbers the vulnerable examples (\ie, positive examples) as shown in~\tab{existing_approach_design_choices}. If left unaddressed, this introduces an undesirable bias in the model limiting its predictive performance. Further, extracted feature vectors of the vulnerable and non-vulnerable examples exhibit a significant overlap in the feature space. This makes it difficult to demarcate the vulnerable examples from the non-vulnerable ones. Training a DL model without accounting for the overlap makes it susceptible to poor predictive performance. 

To mitigate the above problems, we propose a two step approach. First, we use re-sampling to balance the ratio of vulnerable and non-vulnerable examples in the training data. Next, we train a representation learning model on the re-balanced data to learn a representation that can most optimally distinguish vulnerable and non-vulnerable examples. 

%


\subsubsection{Reducing Class Imbalance} 
\label{sect:smote}
In order to handle imbalance in the number of vulnerable and non-vulnerable classes, we use the ``synthetic minority over-sampling technique'' (for short, SMOTE)~\cite{chawla2002smote}. It operates by changing the frequency of the different classes in the data. Specifically, SMOTE sub-samples the majority class (i.e., randomly deleting some examples) while super-sampling the minority class (by creating synthetic examples) until all classes have the same frequency. In the case of vulnerability prediction, the minority class is usually the vulnerable examples. SMOTE has shown to be effective in a number of domains with imbalanced datasets~\cite{tan2015online, lusa2013smote,krawczyk2016learning, agrawal2018better, kellenberger2018detecting, ding2018kernel, tyagi2020sampling}.

\begin{figure}[tp!]
    \algosmote
\end{figure}

During super-sampling, SMOTE picks a vulnerable example and finds $k$ nearest vulnerable neighbors. It then builds a synthetic member of the minority class by interpolating between itself and one of its random nearest neighbors. During under-sampling, SMOTE randomly removes non-vulnerable examples from the training set. This process is repeated until a balance is reached between the vulnerable and non-vulnerable examples. We present the pseudo-code of SMOTE in Algorithm~\ref{alg:smote}.

\subsubsection{Representation Learning Model}
\label{sect:representation_learning}%
The graph embedding of the vulnerable and non-vulnerable code samples at the end of Phase-I tend to exhibit a high degree of overlap in feature space. This effect is illustrated by the t-SNE plot~\cite{maaten2008visualizing} of the feature space in~\fig{t-sne-plots}(a)--(d). In these examples, there are no clear distinctions between the vulnerable  (denoted by \red{$\mathbf{+}$}) and the non-vulnerable samples (denoted by \green{$\mathbf{\circ}$}). This lack of separation makes it particularly difficult to train an ML model to learn the distinction between the vulnerable and the non-vulnerable samples. 

To improve the predictive performance, we seek a model that can project the features from the original non-separable space into a latent space which offers a better separability between vulnerable and non-vulnerable samples. For this, we use a multi-layer perceptron (MLP)~\cite{suter1990multilayer}, designed to transform input feature vector ($x_g$) to a latent representation denoted by $h(x_g)$. The MLP consists of three groups of layers namely, the input layer ($x_g$), a set of intermediate layers which are parameterized by $\theta$ (denoted by $f(\cdot, \theta)$, and a final output layer denoted by $\hat{y}$.

The proposed representation learner works by taking as input the original graph embedding $x_g$ and passing it through the intermediate layers $f(\cdot, \theta)$. The intermediate layer project the original graph embedding $x_g$ onto a latent space $h(x_g)$. Finally, the output layer uses the features in the latent space to predict for vulnerabilities as, $\hat{y} = \sigma\left(\mathrm{W}*h(x_g) + b\right)$.
Where $\sigma$ represents the softmax function, $h_g$ is the latent representation, $\mathrm{W}$ and $b$ represent the model weights and bias respectively.

To maximize the separation between the vulnerable and the non-vulnerable examples in the latent space, we adopt the triplet loss~\cite{mao2019metric} as our loss function. Triplet loss has been widely used in machine learning, specifically in representation learning, to create a maximal separation between classes~\cite{hoffer2015deep, wang2017deep}. The triplet loss is comprised of three individual loss functions: (a)~cross entropy loss ($\mathcal{L}_{CE}$); (b)~projection loss ($\mathcal{L}_{p}$); and (c)~regularization loss ($\mathcal{L}_{\mathit{reg}}$). It is given by:
\begin{equation}
    \mathcal{L}_{trp} = \mathcal{L}_{CE} + \alpha*\mathcal{L}_{p} + \beta*\mathcal{L}_\mathit{reg}
    \label{eqn:full_loss}
\end{equation}
$\alpha$ and $\beta$ are two hyperparameters indicating the contribution of projection loss and regularization loss respectively. The first component of the triplet loss is to measure the cross-entropy loss to  penalize miss-classifications. Cross-entropy loss increases as the predicted probability diverges from the actual label. It is given by, 

\begin{equation}
    \mathcal{L}_{CE} = -\sum\hat{y}\cdot log(y) + (1-\hat{y})\cdot log(1-y)
\end{equation}
Here, $y$ is the true label and $\hat{y}$ represents the predicted label.
The second component of the triplet loss is used the quantify how well the latent representation can separate the vulnerable and non-vulnerable examples. A latent representation is considered useful if all the vulnerable examples in the latent space are close to each other while simultaneous being farther away from all the non-vulnerable examples, \ie, examples from same class are very close (\ie, similar) to each other and examples from different class are far away from each other. Accordingly, we define a loss function $\mathcal{L}_{p}$ which is defined by.

\begin{equation}
     \label{eqn:dist_loss}
     \mathcal{L}_{p} = \left|\mathbb{D}(h(x_g), h(x_\mathit{same})) - \mathbb{D}(h(x_g), h(x_\mathit{diff})) + \gamma\right|
\end{equation}

Here, $h(x_\mathit{same})$ is the latent representation of an example that belongs to the same class as $x_g$ and $h(x_\mathit{diff})$ is the latent representation of an example that belongs to a different class as that of $x_g$. Further, $\gamma$ is a hyperparameter used to define a minimum separation boundary. Lastly, $\mathbb{D(\cdot)}$ represents the cosine distance between two vectors and is given by,

\begin{equation}
    \mathbb{D}(v_1, v_2) = 1 - \left|\frac{v_1 . v_2}{||v_1|| * ||v_2||}\right|
\end{equation}

If the distance between two examples that belong to the same class is large (\ie, $\mathbb{D}(h(x_g), h(x_\mathit{same}))$ is large) or if the distance between two examples that belong to different classes is small (\ie, $\mathbb{D}(h(x_g), h(x_\mathit{diff}))$ is small), $\mathcal{L}_{p}$ would be large to indicate a sub-optimal representation. 

The final component of the triplet loss is the regularization loss ($\mathcal{L}_\mathit{reg}$) that is used to limit the magnitude of latent representation ($h(x_g)$). It has been observed that, over several iterations, the latent representation $h(x_g)$ of the input $x_g$ tend to increase in magnitude arbitrarily~\cite{schroff2015facenet, mao2019metric}. Such arbitrary increase in $h(x_g)$ prevents the model from converging~\cite{girosi1995regularization, pereyra2017regularizing}. Therefore, we use a regularization loss ($\mathcal{L}_\mathit{reg}$) to penalize 
latent representations ($h(x_g)$) that are larger in magnitude. The regularization loss is given by: 

\begin{equation} \mathcal{L}_\mathit{reg} = ||h(x_g)|| + ||h(x_\mathit{same})|| + ||h(x_\mathit{diff})|| \end{equation}

\begin{figure}[tp!]
    \algolossfunc
\end{figure}

With the triplet loss function, \tool trains the model to optimize for it parameters (\ie, $\theta, W, b$) by minimizing equation~\ref{eqn:full_loss}. The effect of using representation learning can be observed by the better separability of the vulnerable and non-vulnerable examples in~\Cref{fig:t-sne-plots}(b). Algorithm~\ref{alg:loss_function} shows the detailed algorithm for calculating the loss.

\section{Experimental Setup}
\label{sec:method}

\subsection{Implementation Details}
\label{subsec:impl_details}
We use Pytorch 1.4.0 with Cuda version 10.1 to implement our method. For GGNN, we use tensorflow 1.15. We ran our experiments on single Nvidia Geforce 1080Ti GPU, Intel(R) Xeon(R) 2.60GHz 16 CPU with 252 GB ram. Neither Devign's implementation, nor their hyperparameters ate not publicly availavle. We followed their paper and re-implemented to our best ability. For the GGNN, maximum iteration number is set to be 500. For the representation learner maximum iteration is 100. We stop the training procedure if F1-score on validation set does not increase in for 50 consecutive training iteration for GGNN and 5 for Representation Learning. Hyper-parameters for different components in \tool are shown in~\Cref{tab:hyperparams}.

\begin{table}[!tpb]
    \scriptsize
    \centering
    \caption{\small Hyper-parameter settings of \tool.}
    \begin{tabular}{l|l|r}
        \hlineB{2}
        \textbf{Model} & \textbf{Parameter} & \textbf{Value} \bigstrut\\
        \hlineB{2}
        \multirow{2}{*}{Word2Vec} & Window Size & 10 \bigstrut\\
        \cline{2-3}
        & Vector Size & 100 \bigstrut\\
        \hlineB{2}
        & Input Embedding Size &  169 \bigstrut\\
        \cline{2-3}
        & Hidden Size & 200 \bigstrut\\
        \cline{2-3}
        & Number of Graph layers & 8 \bigstrut\\
        \cline{2-3}
        & Graph Activation Function & tanh \bigstrut\\
        \cline{2-3}
        \multirow{-7}{*}{GGNN}& Learning Rate & 0.0001\bigstrut\\
        \hlineB{2}
        & Number of hidden layers & 3 \bigstrut\\
        \cline{2-3}
        &Hidden layers sizes & 256, 128, 256\bigstrut\\
        \cline{2-3}
        &Hidden layer activation functions & relu\bigstrut\\
        \cline{2-3}
        &Dropout Probability & 0.2\bigstrut\\
        \cline{2-3}
        & $\gamma$ & 0.5\bigstrut\\
        \cline{2-3}
        & $\alpha$ & 0.5\bigstrut\\
        \cline{2-3}
        &$\beta$ & 0.001\bigstrut\\
        \cline{2-3}
        & Optimizer & Adam\bigstrut\\
        \cline{2-3}
        \multirow{-10}{*}{Repr-model} & $lr$ & 0.001\bigstrut\\
        \hlineB{2}
    \end{tabular}
    \\Repr-model = Representation Learning Model used in \tool.
    \label{tab:hyperparams}
\end{table}


\subsection{Study Subject}
\label{subsec:study_subject}
\Cref{tab:existing_approach_design_choices} summarizes all the \vp approaches and datasets studied in this paper. We evaluate the existing methods (\ie, VulDeePecker~\cite{li2018vuldeepecker}, SySeVR~\cite{li2018sysevr}, Russell~\etal~\cite{russell2018automated}, and Devign~\cite{zhou2019devign}) and \tool's performance on two real world datasets (\ie, \realdata, and \devigndata).  \devigndata was shared by Zhou \etal~\cite{zhou2019devign} who also proposed the \devign model in the same work. Their implementation of \devign was not publicly available. We re-implement their method to report our results. We ensure that our results closely match their reported results in identical settings.

\subsection{Evaluation}
\label{subsec:real-world-evaluation}
To understand a model's  performance, researchers and model developers need to understand the performance of a model against a known set of examples. There are two important aspect to note here, (a) the evaluation metric, and (b) the evaluation procedure. 

\smallskip
\noindent\textit{Problem Formulation and Evaluation Metric:}~
Most of the approaches formulate the problem as a classification problem, where given a code example, the model will provide a binary prediction indicating whether the code is vulnerable or not. This prediction formulation relies on the fact that there are sufficient number of examples (both vulnerable and non-vulnerable) to train on. In this study, we are focusing on the similar formulation. While both VulDeePecker and SySeVR formulate the problem as classification of code slices, we followed the problem formulation used by Russell~\etal~\cite{russell2018automated}, and Devign~\cite{zhou2017automated}, where we classify the function. This is the most suitable model working with the graph, since slices are paths in the graph. 
 
We study approaches based on four popular evaluation metrics for classification task~\cite{powers2011evaluation} -- Accuracy, Precision, Recall, and F1-score. Precision, also known as Positive Predictive rate, is calculated as \textit{true positive / (true positive + false positive)}, indicates correctness of predicted vulnerable samples. Recall, on the other hand, indicates the effectiveness of vulnerability prediction and is calculated as \textit{true positive / (true positive + false negative)}. F1-score is defined as the geometric mean of precision and recall and indicates balance between those.

\smallskip
\noindent\textit{Evaluation Procedure:}~
Since DL models highly depend on the randomness~\cite{beale1996neural}, to remove any bias created due to the randomness, we run 30 trials of the same experiment. At every run, we randomly split the dataset into disjoint  train, validation, and test sets with 80\%, 10\%, and 20\% of the dataset respectively. We report the median performance and the inter-quartile range (IQR) of the performance. When comparing the results to baselines, we use statistical significance test~\cite{koehn2004statistical, smucker2007comparison, agrawal2018better} and effect size test~\cite{hess2004robust}. Significance test tells us whether two series of samples differ merely by random noises. Effect sizes tells us whether two series of samples differ by more than just a trivial amount. 
To assert statistically sound comparisons, following previous approaches~\cite{ferreira2014sisvar,ghotra2015revisiting, mittas2012ranking, canteri2001sasm, jelihovschi2014scottknott}, we use a non-parametric bootstrap hypothesis test~\cite{johnson2001introduction} in conjunction with the A12 effect size test~\cite{10.1145/1985793.1985795, grissom2005effect, vargha2000critique}. We distinguish results from different experiments if both significance test and effect size test agreed that the division was statistically significant (99\% confidence) and is not result of a ``small'' effect (A12 $\ge$ 60\%) (similar to Agrawal~\etal~\cite{agrawal2018better}).


\begin{table}[t!]
\scriptsize
\caption{\small {Performance of existing approaches in predicting real world vulnerability. }
All the numbers are reported as \textit{Median (IQR)} format.}
\label{tab:existing_approaches}
\begin{subfigure}{\linewidth}
\centering
\caption{\small Baseline scores reported by the respective papers. We report single values since authors do not report Median \textit{(IQR)}.}
\label{tab:baseline}
\resizebox{\linewidth}{!}{
\begin{tabular}{@{}l|l|l|rrrr@{}}
\hlineB{2}
\textbf{Dataset} & \textbf{Technique} & \textbf{Training}  & \textbf{Acc} & \textbf{Prec} & \textbf{Recall}   & \textbf{F1}\bigstrut\\\hlineB{2}
 & VulDeePecker & NVD/SARD  & $\cdot$ & 86.90 & $\cdot$ & 85.40\bigstrut\\
\cline{2-7}
 & SySeVR     & NVD/SARD  & 95.90 & 82.50 & $\cdot$ & 85.20\bigstrut\\
 \cline{2-7}
 & \multirow{2}{*}{Russell~\etal} & Juliet  & $\cdot$ & $\cdot$ & $\cdot$ & 84.00\bigstrut\\
  \cline{3-7}
\multirow{-6}{*}{\rotatebox{90}{Baseline}} & & \draper      
                & $\cdot$ & $\cdot$ & $\cdot$ & 56.6\bigstrut\\
\cline{2-7}
& Devign & \devigndata      
                & 72.26 & $\cdot$ & $\cdot$ & 73.26\bigstrut\\
\hlineB{2}
\multicolumn{7}{r}{$\cdot$ = Not Reported.}\bigstrut\\
\end{tabular}}
\end{subfigure}
~
\begin{subfigure}{\linewidth}
\centering
\caption{\small Scenario-A: Using Existing Pre-trained Models}
\label{tab:pretrained}
\resizebox{\linewidth}{!}{
\begin{tabular}{@{}l|l|l|rrrr@{}}
\hlineB{2}
\textbf{Dataset} & \textbf{Technique} & \textbf{Training}  & \textbf{Acc} & \textbf{Prec} & \textbf{Recall}   & \textbf{F1}\bigstrut\\ 
\hlineB{2}
& VulDeePecker & NVD/SARD  & 79.05 & 11.12 & 13.64 & 12.18\bigstrut[t]\\
& & & \textit{(0.25)} & \textit{(0.48)} & \textit{(0.50)} & \textit{(0.47)} \bigstrut[b]\\
\cline{2-7}
 & SySeVR & NVD/SARD  & 79.48 & 9.38  &  15.89  & 10.37 \bigstrut[t]\\
 & & & \textit{(0.24)} & \textit{(0.30)} & \textit{(0.63)} & \textit{(0.36)} \bigstrut[b]\\
 \cline{2-7}
 & \multirow{2}{*}{Russell~\etal} & Juliet &  38.11  & 41.36  & 6.51 & 11.24 \bigstrut[t]\\
 & & & \textit{(0.11)} & \textit{(0.38)} & \textit{(0.07)} & \textit{(0.12)} \bigstrut[b]\\
 \cline{3-7}
 & & \draper     
          & 70.08  & 49.05 & 15.61 & 23.66 \bigstrut[t]\\
          & & & \textit{(0.14)} & \textit{(0.35)} & \textit{(0.12)} & \textit{(0.24)} \bigstrut[b]\\
    \cline{2-7}
  \multirow{-10}{*}{\rotatebox{90}{\tool}~~\rotatebox{90}{~dataset}} & \multirow{2}{*}{Devign} & \multirow{2}{*}{\devigndata} &   66.24 & 10.74 & 37.04 & 16.68 \bigstrut[t]\\
        & & & \textit{(0.14)} & \textit{(0.11)} & \textit{(0.54)} & \textit{(0.17)}\bigstrut[b]\\
\hlineB{2}
 & VulDeePecker &  NVD/SARD  & 52.27 & 8.51 & 44.78 & 14.27 \bigstrut[t]\\
 & & & \textit{(0.23)} & \textit{(0.22)} & \textit{(0.66)} & \textit{(0.33)} \bigstrut[b]\\
\cline{2-7}
 & SySeVR &  NVD/SARD  & 52.52 & 10.62 &  46.69 & 16.77 \bigstrut[t]\\
 & & & \textit{(0.18)} & \textit{(0.22)} & \textit{(0.20)} & \textit{(0.31)} \bigstrut[b]\\
 \cline{2-7}
 & \multirow{4}{*}{Russell~\etal} & Juliet &  49.84 & 33.17 & 45.53 & 37.65 \bigstrut[t]\\
 & & & \textit{(0.10)} & \textit{(0.13)} & \textit{(0.14)} & \textit{(0.12)} \bigstrut[b]\\
  \cline{3-7}
\multirow{-8}{*}{\rotatebox{90}{FFMpeg +}~~\rotatebox{90}{~Qemu}} &  & \draper      
       & 53.96 & 44.00 & 49.53 & 46.60 \bigstrut[t]\\
       & & & \textit{(0.14)} & \textit{(0.17)} & \textit{(0.20)} & \textit{(0.15)} \bigstrut[b]\\
\hlineB{2}
\end{tabular}
}
\end{subfigure}

\vspace{3mm}
\begin{subfigure}{0.99\linewidth}
\centering
\caption{\small Scenario-B: Using Retrained Models with Real-world Data.}
    \label{tab:retrained}
    \footnotesize
    \resizebox{\linewidth}{!}{
    \begin{tabular}{@{}l|l|l|rrrr@{}}
        \hlineB{2}
        \textbf{Dataset} & \textbf{Input} & \textbf{Approach} &  \textbf{Acc} & \textbf{Prec}  & \textbf{Recall} & \multicolumn{1}{c}{\textbf{F1}}\bigstrut\\
        \hlineB{2}
        \multirow{8}{*}{\rotatebox{90}{\tool}~~\rotatebox{90}{~dataset}} & \multirow{1}{*}{Token} &
		Russell~\etal &  {{90.98}} & 24.63  & 10.91  & 15.24 \bigstrut[t]\\
		& & & \textit{({0.75})} & \textit{(5.35)} & \textit{(2.47)} & \textit{(2.74)} \bigstrut[b]\\
        \cline{2-7}
        & \multirow{2}{*}{Slice~+~} &
		VulDeePecker &  89.05  & 17.68  & 13.87  & 15.7    \bigstrut[t]\\
		& & & \textit{(0.80)} & \textit{(7.51)} & \textit{(8.53)} & \textit{(6.41)} \bigstrut[b]\\
        \cline{3-7}
        & \multirow{-2}{*}{Token} &
        SySeVR &  84.22  & 24.46  & {40.11} & {30.25} \bigstrut[t]\\	
        & & & \textit{(2.48)} & \textit{(4.85)} & \textit{({4.71})} & \textit{({2.35})} \bigstrut[b]\\
        \cline{2-7}
        & \multirow{1}{*}{Graph} &
        Devign &  88.41 & {{34.61}} & 26.67 & 29.87  \bigstrut[t]\\
        & & & \textit{(0.66)} & \textit{({3.24})} & \textit{(6.01)} & \textit{(4.34)} \bigstrut[b]\\

        
      \hlineB{2}
       
        \multirow{8}{*}{\rotatebox{90}{FFMpeg +}~~\rotatebox{90}{~Qemu}} & \multirow{1}{*}{Token} &
		Russell~\etal &  58.13 & {54.04} & 39.50 & 45.62  \bigstrut[t]\\
		& & & \textit{(0.88)} & \textit{({2.09})} & \textit{(2.17)} & \textit{(1.33)} \bigstrut[b]\\
        \cline{2-7}
        & \multirow{2}{*}{Slice~+~} &
		VulDeePecker & 53.58 & 47.36 & 28.70 & 35.20 \bigstrut[t]\\
		& & & \textit{(0.61)} & \textit{(1.80)} & \textit{(12.08)} & \textit{(8.82)} \bigstrut[b]\\
        \cline{3-7}
        & \multirow{-2}{*}{Token} &
        SySeVR &  52.52 & 48.34 & {65.96} & 56.03 \bigstrut[t]\\
        & & & \textit{(0.81)} & \textit{(1.51)} & \textit{({7.12})} & \textit{(3.20)} \bigstrut[b]\\
        \cline{2-7}
        & \multirow{1}{*}{Graph} &
        Devign$^\dagger$ &  {58.57} & 53.60 & 62.73 & {57.18} \bigstrut[t]\\	
        & & & \textit{({1.03})} & \textit{(3.21)} & \textit{(2.99)} & \textit{({2.58})} \bigstrut[b]\\

      \hlineB{2}
    \end{tabular}
    }
    \begin{flushleft}
    {$^\dagger$ We made several unsuccessful attempts to contact the authors for Devign's implementation. Despite our best effort, Devign's reported result is not reproducible. We make our implementation of Devign public at  \href{https://github.com/saikat107/Devign}{\it https://github.com/saikat107/Devign} for further use.}
    \end{flushleft}
    \end{subfigure}
\end{table}

\section{Empirical Results}
\label{sec:results}
We present our empirical results as answers to the following research questions:
\bi
\item \textbf{RQ1:} {How effective are existing approaches for real-world \vp?} (\tion{rq1})
\item \textbf{RQ2:} {What are the limitations of existing approaches?} (\tion{rq2})
\item \textbf{RQ3:} {How to improve \dlvp approaches?} (\tion{rq3})

\ei


\subsection{Effectiveness of existing \vp approaches (RQ1)}
\label{sect:rq1}
\motivation The goal of any \dlvp approaches is to be able to predict vulnerabilities in the real-world. The datasets that the existing models are trained on contain simplistic examples that are representative of real-world vulnerabilities. Therefore, we ought to, in theory, be able to use these models to detect vulnerabilities in the real-world. 

\approach 
There are two possible scenarios under which these models may be used: 

\noindent$\circ$~ \textit{Scenario-A~(pre-trained models)}:~
We may reuse the existing pre-trained models as it is to predict real-world vulnerabilities. To determine how they perform in such a setting, we first train the baseline models with their respective datasets as per~\Cref{tab:existing_approach_design_choices}.  Next, we use those pre-trained models to detect vulnerabilities in the real-world (\ie, on \devigndata, and \tool dataset).


\noindent$\circ$~ \textit{Scenario-B~(re-trained models)}:~
We may rebuild the existing models first by training them on the real-world datasets, and then use those models to detect the vulnerabilities. To assess the performance of baseline approaches in this setting, we first use one portion of the \devigndata and \tool dataset to train each model. Then, we use those models to predict for vulnerabilities in the remainder of the \devigndata and \tool. We repeat the process 30 times, each time training and testing on different portions of the dataset.


\observations 
\tab{pretrained} tabulates the performance of existing pre-trained models on predicting vulnerabilities in real-world data 
(\ie, Scenario-A). We observe a precipitous drop in performance when pre-trained models are used for real-world vulnerability prediction. 

For example, In \realdata, VulDeePecker achieves an F1-score of \textit{only} $\mathit{12.18}\%$ and in \devigndata, VulDeePecker achieves an F1-score of $\mathit{14.27}\%$, while in the baseline case (see~\Cref{tab:baseline}), the F1-score of VulDeePecker was as high as $\mathit{85.4\%}$. Even the sophisticated graph-based Devign model produced an F1-score of only $\sim17\%$ and precision as low as $\sim10\%$ on \realdata. 
Similar performance drops are observed for all the other baselines. 
On average, we observe a 73\% drop of F1-score across all the models in this setting.  

For scenario-B, \tab{retrained} tabulates our findings for re-trained models.  Here, we also observe a significant performance drop from the baseline results. In \tool dataset, both Russell~\etal and VulDeePecker achieve an F1-score of roughly $\mathit{15\%}$ (in contrast to their baseline performances of $\mathit{85\%}$). SySeVR achieved an F1-score of $\mathit{30\%}$ on \tool dataset. We observed similar trends in other settings, 
with an average F1 score drop of 54\%.

\begin{result}
Existing approaches fail to generalize to real-world vulnerability prediction. If we directly use a pre-trained model to detect the real-world vulnerabilities, the f1-score drops by $\sim$\textit{73\%}, on average. Even if we retrain these models with real-world data, their performance drops by $\sim$\textit{54\%} from the reported results.
\end{result}


\subsection{Key limitations of existing \dlvp approaches (RQ2)}
\label{sect:rq2}

\motivation
In RQ1, we showed that existing approaches are not effective in detecting real-world vulnerabilities.  In this RQ, we investigate the reasons behind their failure. We find that the baseline methods suffer from a number of problems, as listed below: 

\begin{table}[!tb]
    \centering
    \footnotesize
    \caption{{\small Percentage of duplicate samples in datasets.}}
    \label{tab:duplicate}
    \resizebox{0.9\linewidth}{!}{
        \begin{tabular}{l|l|r}
            \hlineB{2}
            \textbf{Dataset} &\textbf{Pre-processing Technique} & \textbf{\% of duplicates}\bigstrut\\
            \hlineB{2}
             Juliet & Russell~\etal & 68.63 \bigstrut\\
             \hlineB{1}
             \multirow{2}{*}{NVD + SARD} & VulDeePecker &  67.33 \bigstrut\\
             & SySeVR &  61.99 \bigstrut\\
             \hlineB{1}
             \draper & Russell~\etal & 6.07 / 2.99 \bigstrut\\
             \hlineB{1}
             \multirow{4}{*}{\realdata} & None & 0.6\bigstrut\\
             &VulDeePecker & 25.85\bigstrut\\
             & SySeVR & 25.56 \bigstrut\\
             & Russell~\etal & 8.93\bigstrut\\
             \hline
             \multirow{4}{*}{\devigndata} & None & 0.2\bigstrut\\
             & VulDeePecker & 19.58\bigstrut\\
             & SySeVR & 22.10 \bigstrut\\
             & Russell~\etal & 20.54\bigstrut\\
             \hlineB{2}
        \end{tabular}
    }
\vspace{-3mm}
\end{table}

\subsubsection{Data Duplication}
Preprocessing techniques such as slicing used by VulDeePecker and SySeVR and tokenization used by Russell~\etal introduce a large number of duplicates in both the training and testing data. There are several ways duplication can be introduced by these preprocessing techniques -- \eg same slice can be extracted from different entry points, different code can have same tokens due to the abstract tokenization, etc.

\approach 
 We apply each preprocessing technique to its respective dataset (see~\tion{background}) and also to the real-world datasets. 

\observations 
\tab{duplicate} tabulates the number of duplicates introduced by some of the \vp approaches. We observe that the preprocessing technique of  SySeVR and VulDeePecker (\ie, slicing followed by tokenization) introduces a significant amount of ($>\mathit{60\%}$) duplicate samples.  Further,  semi-synthetic datasets like NVD, SARD, and Juliet (comprised of much simpler code snippets) result in a large number of duplicates. In contrast, real-world datasets are much more complex and therefore have far fewer duplicates. In our case, the two real-world data contain little to no duplicates prior to preprocessing (\realdata had only 0.6\%, and \devigndata had 0.2\%). 
After preprocessing, although some duplicates are introduced (\eg SySeVR's preprocessing technique introduces 25.56\% duplicates in \realdata and 22.10\% duplicates in \devigndata), they are much lesser than baseline datasets. While duplicates created by slicing and pre-processing techniques do favor \vp in general~\cite{li2018sysevr, zou2019muvuldeepecker}, it seriously undermines the capability of a DL model to extract patterns. In fact, prevalence of such duplicates in training set might lead a DL model to learn irrelevant features. Common examples between train and test sets hampers fair comparison of different DL models for \vp task.

Ideally, a DL based model should be trained and tested on a dataset where 100\% examples are unique. Duplication tends to artificially inflate the overall performance of a method~\cite{allamanis2019adverse}, as evidenced by the discrepancy of the baseline results and results of the pre-trained models in Scenario-A of RQ1 (see \tab{pretrained}).

\subsubsection{Data Imbalance} 
\label{data-imbalance}
Real world data often contains significantly more non-vulnerable examples than vulnerable ones. A model trained on such skewed dataset is susceptible to being considerably biased toward the majority class. 

\approach We compute percentage on vulnerable samples \wrt total number of samples from different datasets used in this paper as shown in ~\tab{existing_approach_design_choices}.

\observations
We notice that several datasets exhibit a notable imbalance in the fraction of vulnerable and non-vulnerable examples; . 
the percentage vulnerability is sometimes as low as $6\%$. The ratio of vulnerable and non-vulnerable examples varies depending on the project and the data collection strategy employed.
Existing methods fail to adequately address the data imbalance during training. This causes two problems: (1) When pre-trained models are used (\ie, Scenario-A in RQ1) to predict vulnerabilities in the real world, the ratios of vulnerable and non-vulnerable examples differ significantly in training and testing datasets. This explains why pretrained models perform poorly (as seen  in~\tab{pretrained}). (2) When the models are re-trained, they tend to be biased towards the class with the most examples (\ie, the majority class). This results in poor recall values (\ie, they miss a lot of true vulnerabilities) and hence, also the F1-score (as seen in~\tab{retrained}).   
\begin{figure}
    \centering
    \begin{subfigure}{0.95\linewidth}
    \scriptsize
    \tt
    \begin{tabular}{|r p{0.89\linewidth}|}
    \hline
    1  & \ltwo{link\_layer\_show(\textbf{struct} ib\_port {*}p, } \\
    2  & ~~~~~~~\lone{\textbf{struct} port\_attribute {*}unused, \textbf{char} * buf)\{}\\
    3  & ~~~\lone{\textbf{switch} }\lthree{(rdma\_port\_get\_link\_layer(}\\
    4  & ~~~~~~~~~~~~~~~\lthree{p->ibdev, p->port\_num))}\lfive{ \{}\\
    5  & ~~~~~~\lfive{\textbf{case} }\lfour{IB\_LINK\_LAYER\_INFINIBAND:}\\
    6  & ~~~~~~~~~\lfour{\textbf{return} \Red{sprintf(buf, "\%s$\backslash$n", "InfiniBand")};}\\
    7  & ~~~~~~\lfive{\textbf{case} IB\_LINK\_LAYER\_ETHERNET:}\\
    8  & ~~~~~~~~~\lfive{\textbf{return} \Red{sprintf(buf, "\%s$\backslash$n", "Ethernet")};}\\
    9  & ~~~~~~\lfive{\textbf{default}:}\\
    10 & ~~~~~~~~~\lfive{\textbf{return} \Red{sprintf(buf, "\%s$\backslash$n", "Unknown")};}\\
    11 & ~~~\lfive{\}}\\
    12 & \lfive{\}}\\
    \hline
    \end{tabular}
    \caption{{\small Vulnerable code example in \draper \cite{russell2018automated} dataset correctly predicted by Russel~\etal's token-based method.\\}}
    \label{fig:lemna_draper}
    \end{subfigure}
    \quad\quad
    \begin{subfigure}{0.95\linewidth}
    \scriptsize
    \tt
    \begin{tabular}{|r p{0.89\linewidth}|}
    \hline
    1 & \textbf{static int} mov\_read\_dvc1(\lfour{MOVContext *c},  \\
    2 & ~~~~~~~~\lfour{AVIOContext *pb}, \lthree{MOVAtom atom}) \{ \\
    3 & ~~~~\lone{AVStream *st};  \\
    4 & ~~~~\lfour{\textbf{uint8\_t} profile\_level}; \\
    5 & ~~~~\textbf{if} (\lthree{c->fc->nb\_streams < 1}) \\
    6 & ~~~~~~~~\lfour{\textbf{return} 0}; \\
    7 & ~~~~\ltwo{st = c->fc->streams[c->fc->nb\_streams-1]}; \\
    8 & ~~~~\textbf{if} (\lfour{atom.size $>=$ (1$<<$28) || atom.size $<$ 7}) \\
    9 & ~~~~~~~~\lfour{\textbf{return} AVERROR\_INVALIDDATA}; \\
    10 & ~~~~\lfour{profile\_level = avio\_r8(pb)}; \\
    11 & ~~~~\textbf{if} (\lfour{(profile\_level \& 0xf0) $!=$ 0xc0}) \\
    12 & ~~~~~~~~\lthree{\textbf{return} 0}; \\
    ...   & ~~~~...\\
    18 & ~~~~\lfour{st->codec->extradata\_size = atom.size - 7}; \\
    19 & ~~~~\ltwo{avio\_seek(pb, 6, SEEK\_CUR)}; \\
    20 & ~~~~\lfour{\Red{{avio\_read(}}}\\
    21 & ~~~~~~~~\lfour{\Red{{pb, st->codec->extradata,}}}\\
    22 & ~~~~~~~~~~~~\lfour{\Red{{st->codec->extradata\_size)}}};\\
    23 & ~~~~\lfour{\textbf{return} 0}; \\
    24 & \}\\
    \hline
    \end{tabular}
    \caption{{\small Vulnerable example from \devigndata~\cite{zhou2019devign} dataset correctly predicted by graph model. Other method could not predict the vulnerability in this example.}}
    \label{fig:lemna_ggnn}
    \end{subfigure}
    \caption{\small{{Contribution of different code component in correct classification of vulnerability by different model.} \lone{Red-shaded}  code elements are most contributing, \lfive{Green-shaded} are the least. \Red{Red} colored code are the source of vulnerabilities.}}
    \label{fig:code_heatmap}
    \end{figure}

\begin{figure*}[tpb!]
    \centering
    \begin{subfigure}{0.19\linewidth}
    \centering
    \includegraphics[width=\linewidth]{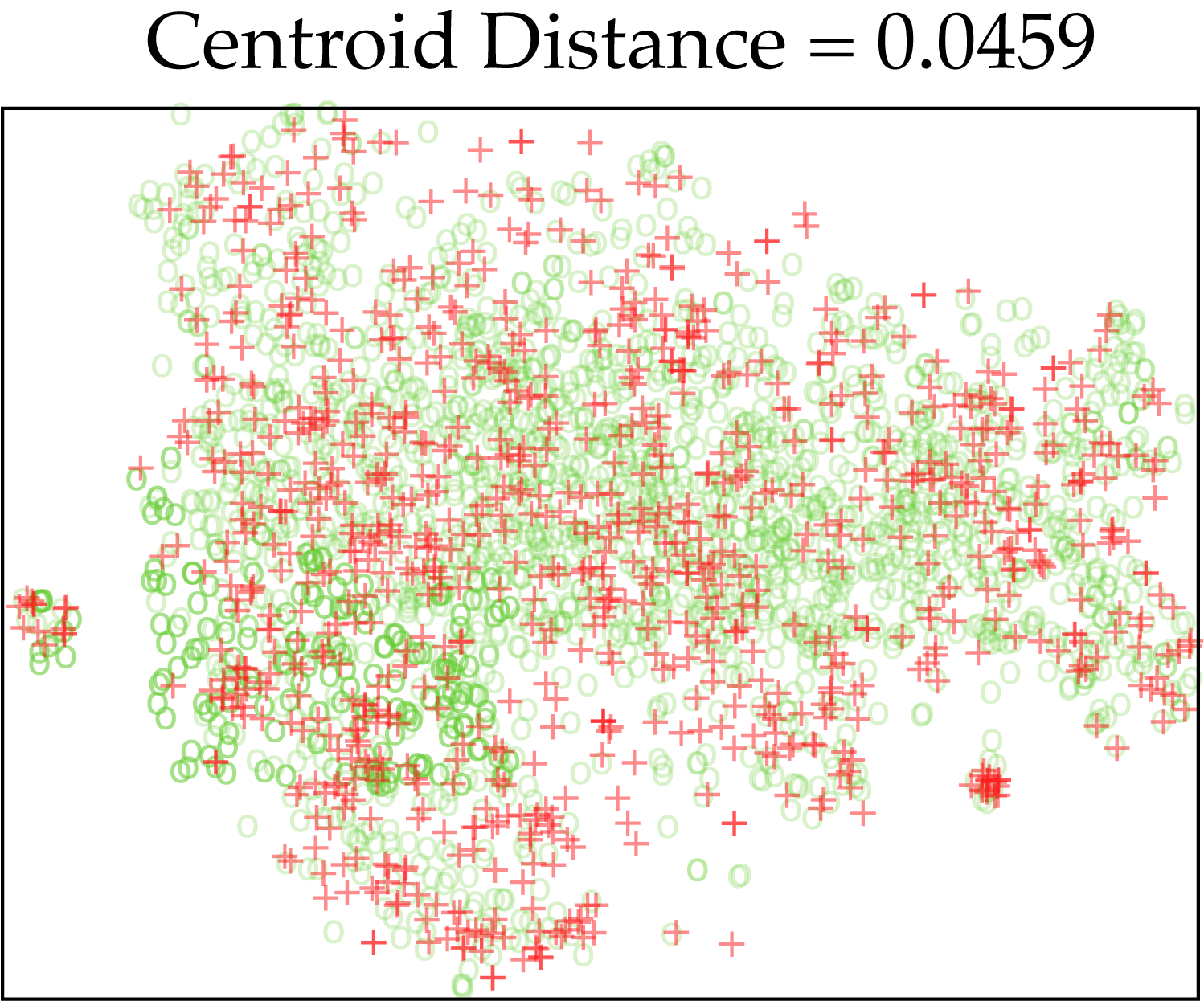}
    \caption{Russell~\etal }
    \label{fig:russell-t-sne}
    \end{subfigure}
    \begin{subfigure}{0.19\linewidth}
    \centering
    \includegraphics[width=\linewidth]{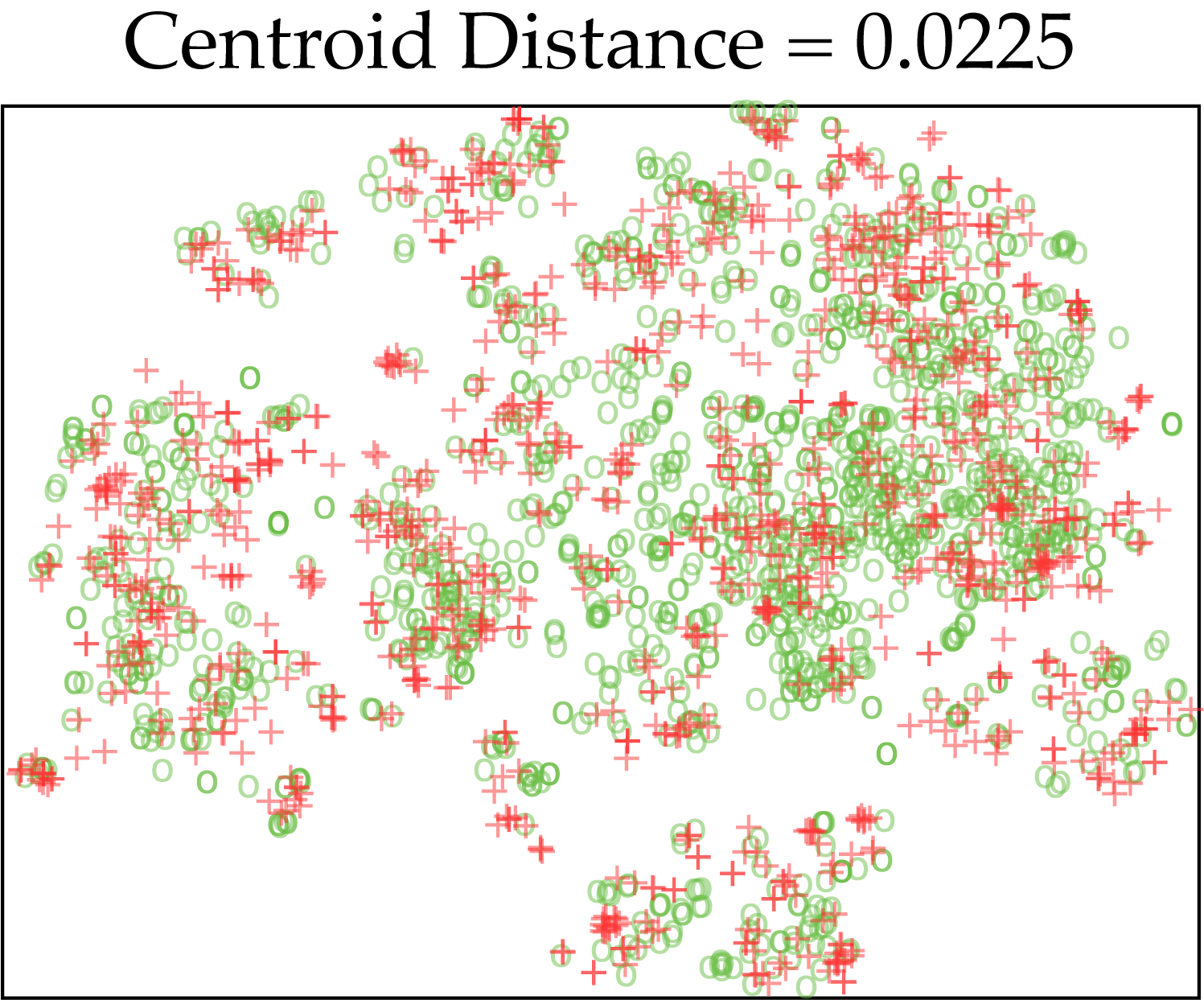}
    \caption{VulDeePecker}
    \label{fig:vuldeepecker-t-sne}
    \end{subfigure}
    \begin{subfigure}{0.19\linewidth}
    \centering
    \includegraphics[width=\linewidth]{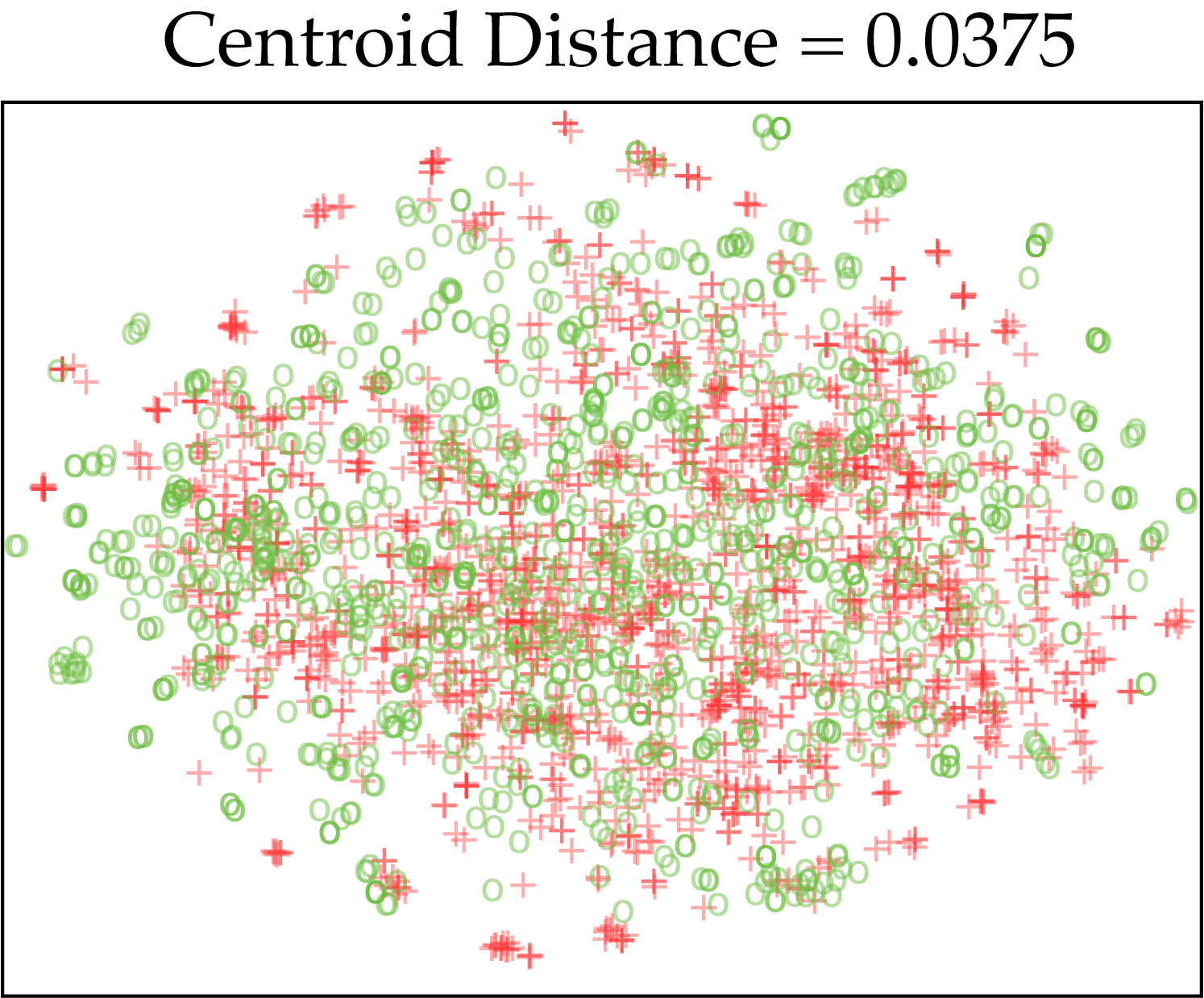}
    \caption{SySeVR}
    \label{fig:sysevr-t-sne}
    \end{subfigure}
    \begin{subfigure}{0.19\linewidth}
        \centering
        \includegraphics[width=\linewidth]{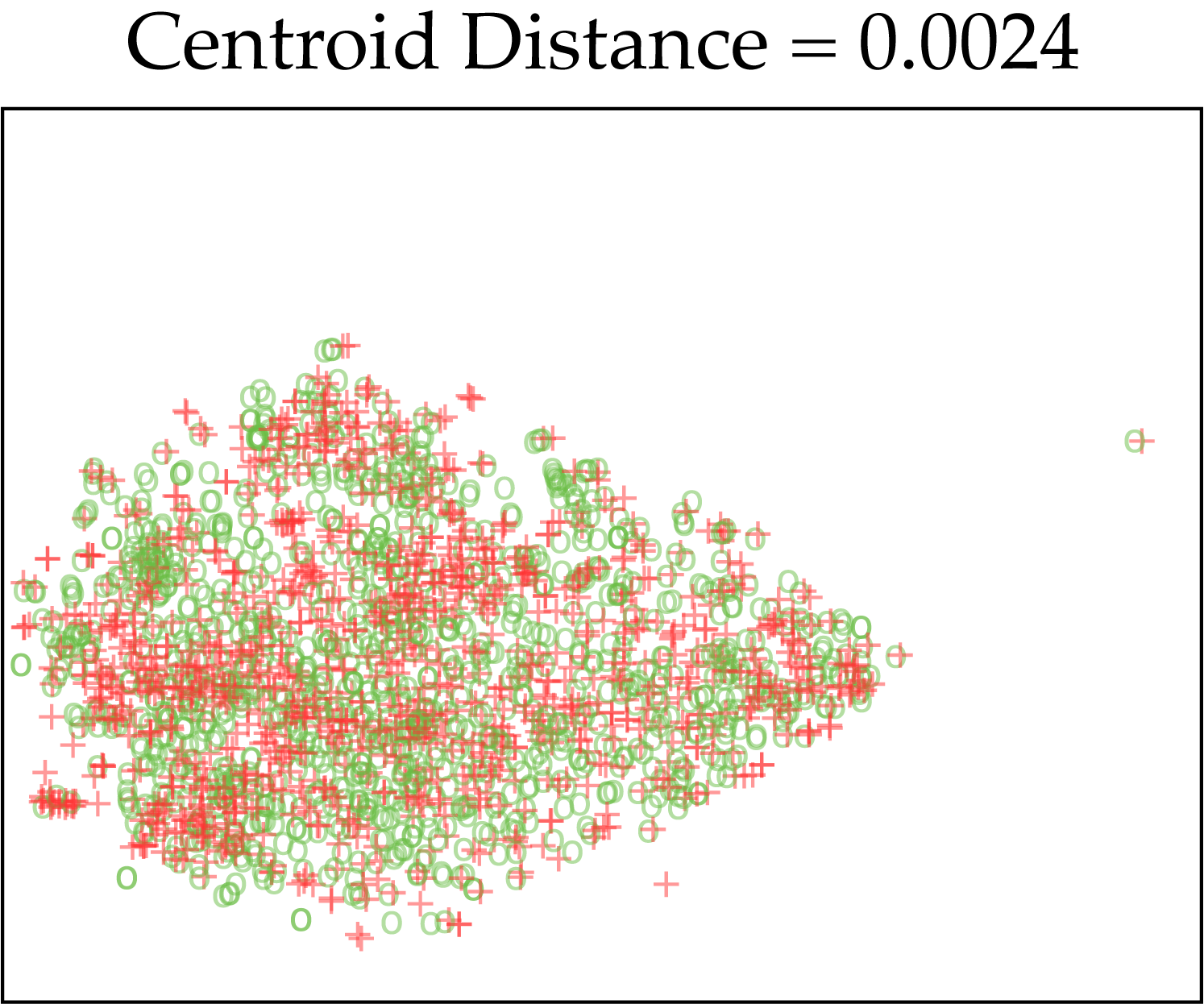}
        \caption{Devign}
        \label{fig:devign-t-sne}
    \end{subfigure}    
    \begin{subfigure}{0.19\linewidth}
    \centering
    \includegraphics[width=\linewidth]{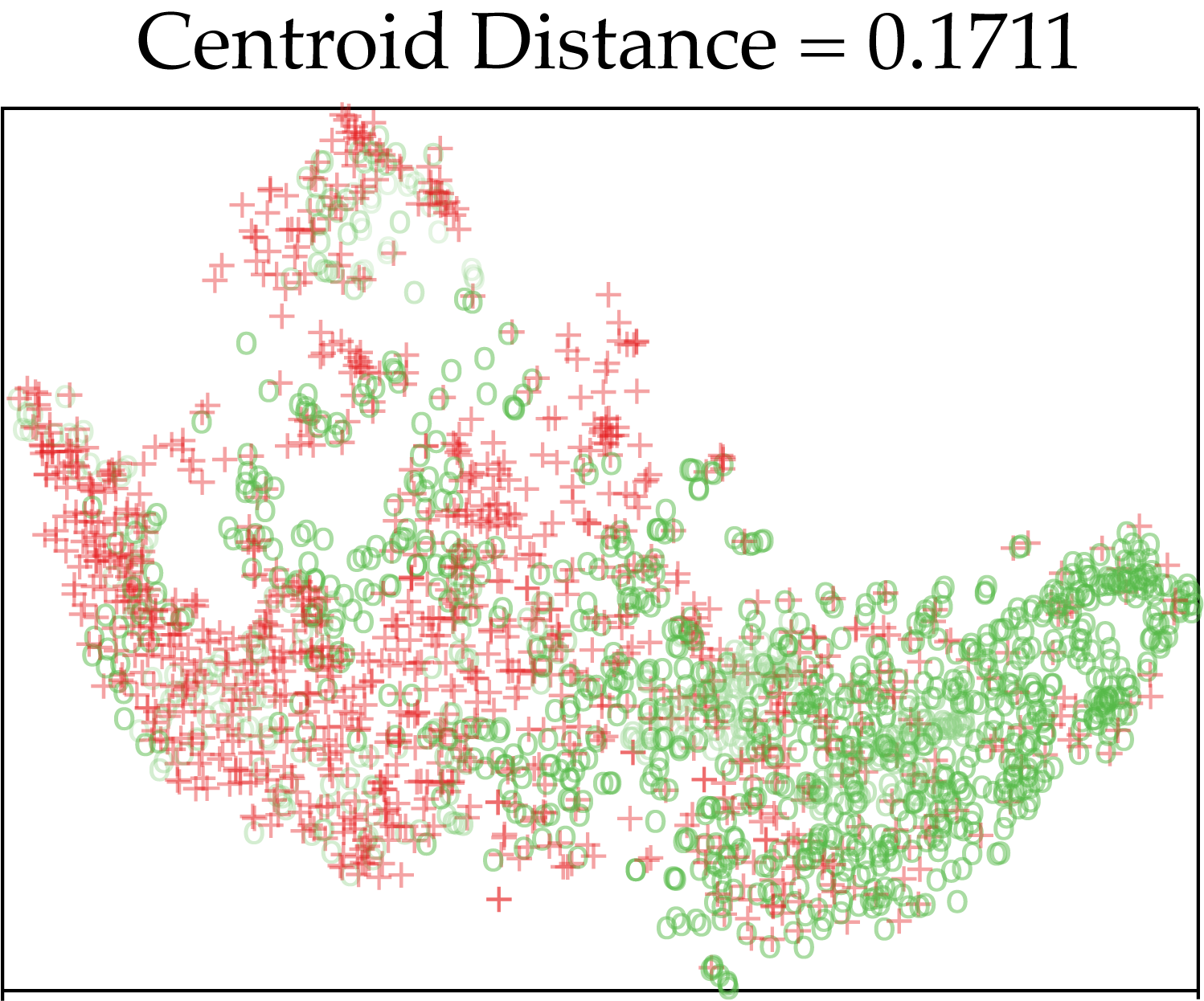}
    \caption{\tool}
    \label{fig:reveal-t-sne}
    \end{subfigure}
    \caption{{\small t-SNE plots illustrating the separation between vulnerable (denoted by \red{$+$}) and non-vulnerable (denoted by {\green{$\mathbf{\circ}$}}) example. Existing methods fail to optimally separate vulnerable and non-vulnerable classes.}}
    \label{fig:t-sne-plots}
\end{figure*}

\noindent\subsubsection{Learning Irrelevant Features}
In order to choose a good DL model for \vp, it is important to understand what features the model uses to make its predictions. A good model should assign greater importance to the vulnerability related code features.

\approach To understand what features a model uses for its prediction, we find the feature importance assigned to the predicted code by the existing approaches. For token-based models such as VulDeePecker, SySeVR, and Russell \etal, we use Lemna to identify feature importance~\cite{guo2018lemna}. Lemna assigns each token in the input with a value $\omega_i^t$, representing the contribution of that token for prediction. A higher value of $\omega_i^t$ indicates a larger contribution of token towards the prediction and vice versa.  
For graph-based models, such as Devign, Lemna is not applicable~\cite{guo2018lemna}. 
In this case, we use the activation value of each vertex in the graph to obtain the feature importance. The larger the activation, the more critical the vertex is. 

\observations To visualize the feature importances, we use a heatmap to highlight the most to least important segments of the code. \fig{code_heatmap} shows two examples of correct predictions. \fig{lemna_draper} shows an instance where Russell \etal's token-based method accurately predicted a vulnerability. But, the features that were considered most important for the prediction (lines 2 and 3) are not related to the actual vulnerability that appears in buggy \texttt{sprintf} lines (lines 6, 8, and 10). We observe similar behavior in other token based methods.

In contrast, \fig{lemna_ggnn} shows an example that was misclassified as non-vulnerable by token-based methods, but graph-based models accurately predict them as vulnerable. Here we note that the vulnerability is on line 20, and graph-based models use lines 3, 7, 19 to make the prediction, \ie mark the corresponding function as vulnerable. We observe that each of these lines shares a data dependency with line 20 (through \texttt{pb} and \texttt{st}). Since graph-based models learn the semantic dependencies between each of the vertices in the graph through the code property graph, a series of connected vertices, each with high feature importance, causes the graph-based model to make the accurate prediction. Token-based models lack the requisite semantic information and therefore fail to make accurate predictions.

\subsubsection{Model Selection: Lack of Class Separation} 

Existing approaches translate source code into a numeric feature vector that can be used to train a \vp model. The efficacy of the \vp model depends on how separable the feature vectors of the two classes (\ie, vulnerable examples and non-vulnerable examples) are. The greater the separability of the classes, the easier it is for a model to distinguish between them. 

\approach We use t-SNE plots to inspect the separability of the existing models. t-SNE is a popular dimensionality reduction technique that is particularly well suited for visualizing how high-dimensional datasets look in a feature space~\cite{maaten2008visualizing}. A clear separation in the t-SNE space indicates that the classes are distinguishable from one another. In order to numerically quantify the separability of the classes, we use the centroid distance proposed by Mao \etal~\cite{mao2019metric}. We first find the centroids of each of the two classes. Next, we compute the euclidean distance between the the centroids. Models that have larger the euclidean distances are preferable since they exhibit greater class separation. 

\observations \fig{t-sne-plots} illustrates the t-SNE plots of the existing approaches. All the existing approaches (\fig{russell-t-sne}--\ref{fig:devign-t-sne}) exhibit a significant degree of overlap in the feature space between the two classes. This is also reflected by the relatively low distance between the centroids in each of the existing methods. Among exiting methods, Devign (\fig{devign-t-sne}) has the least centroid distance (around $0.0025$); this is much lower than any other existing approach. This lack of separation explains why Devign, in spite of being a graph-based model, has poor real-world performance (see~\tab{existing_approaches}).

\begin{result}
Existing approaches have several limitations: they (a) introduce data duplication, (b) don't handle data imbalance, (c) don't learn semantic information, (d) lack class separability. \dlvp may be improved by addressing these limitations.
\end{result}

\begin{figure*}[!htb]
    \centering
    \begin{minipage}[t]{0.49\textwidth}
        \begin{subfigure}{0.30\linewidth}
        \includegraphics[width=\linewidth]{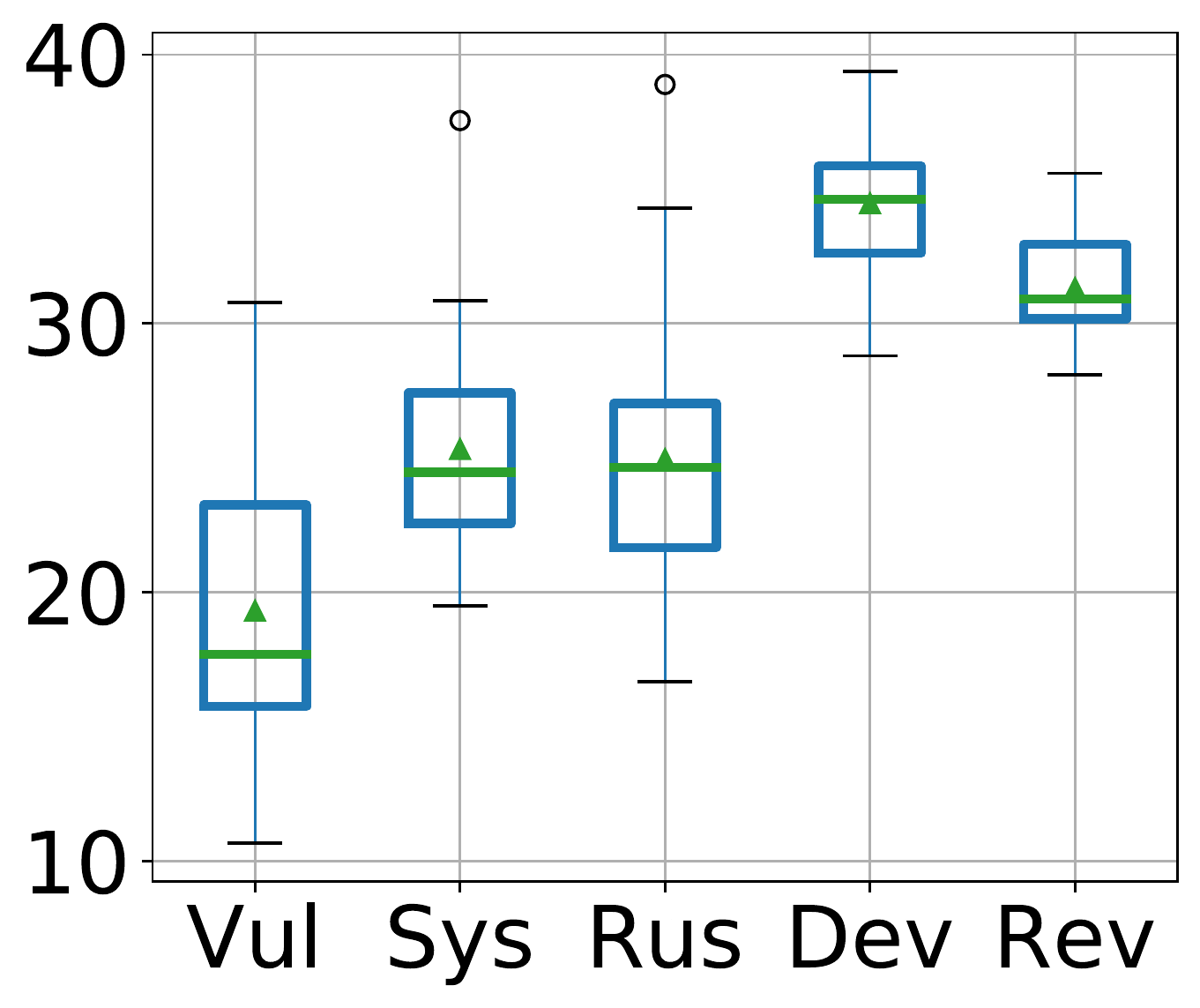}
        \caption{\small Precision}
        \label{fig:verun_pr_bp}
        \end{subfigure}
        \begin{subfigure}{0.30\linewidth}
        \includegraphics[width=\linewidth]{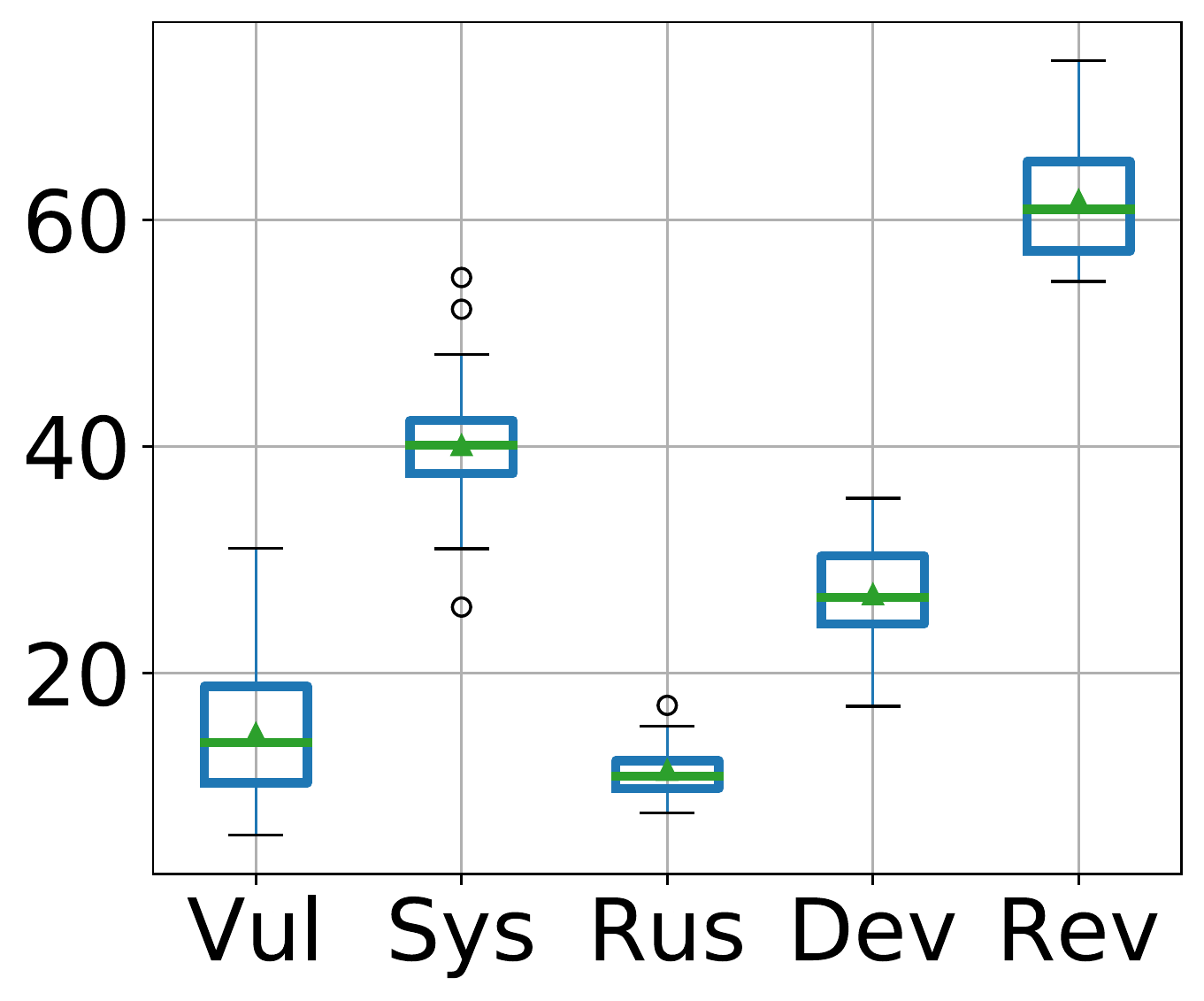}
        \caption{\small Recall}
        \label{fig:verun_rc_bp}
        \end{subfigure}
        \begin{subfigure}{0.30\linewidth}
        \includegraphics[width=\linewidth]{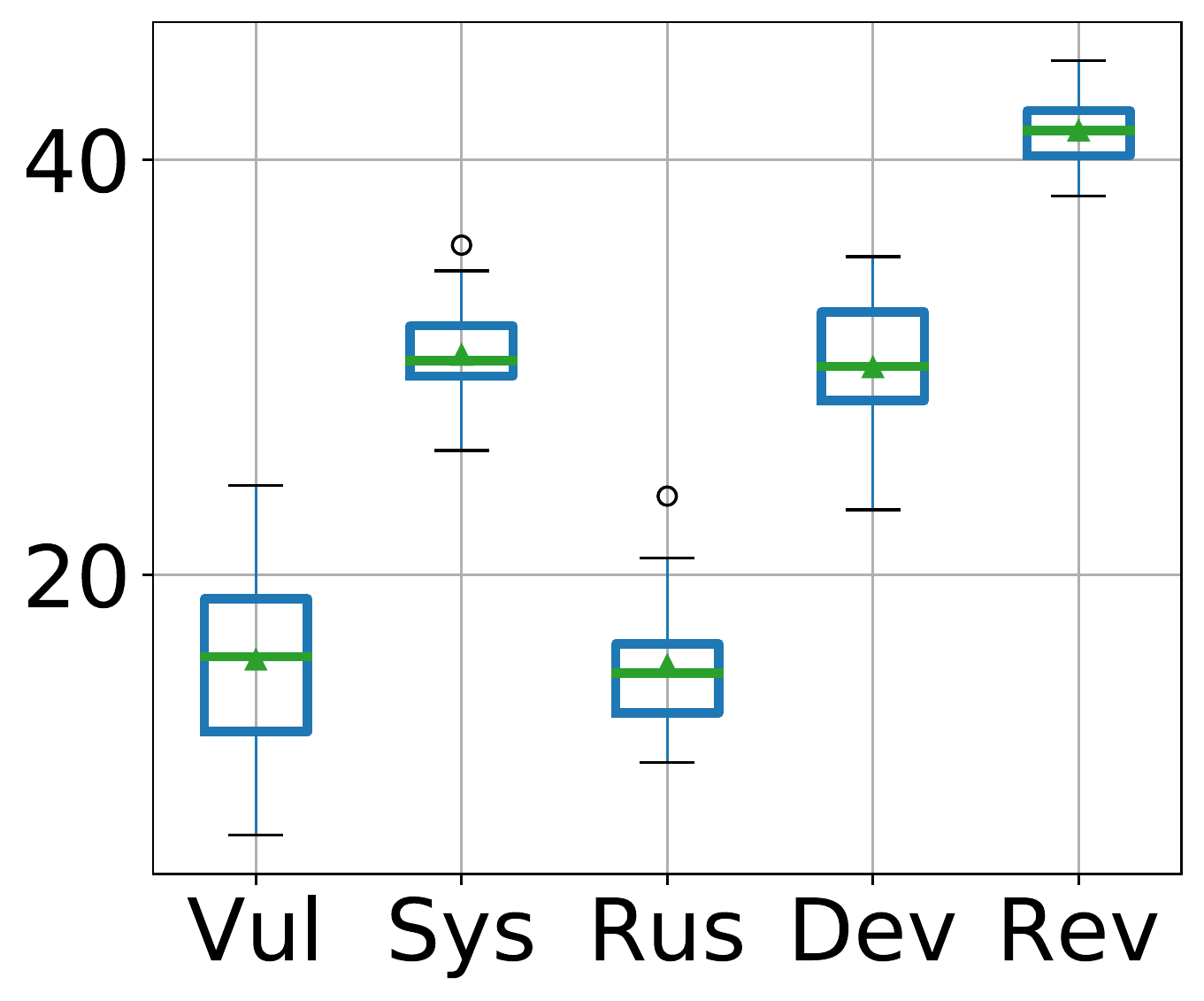}
        \caption{\small F1-score}
        \label{fig:verun_f1_bp}
        \end{subfigure}
        \caption{\small {Performance spectrum of \realdata.}}
        \label{fig:verun_bp}
    \end{minipage}%
    \begin{minipage}[t]{0.49\textwidth}
        \begin{subfigure}{0.30\linewidth}
        \includegraphics[width=\linewidth]{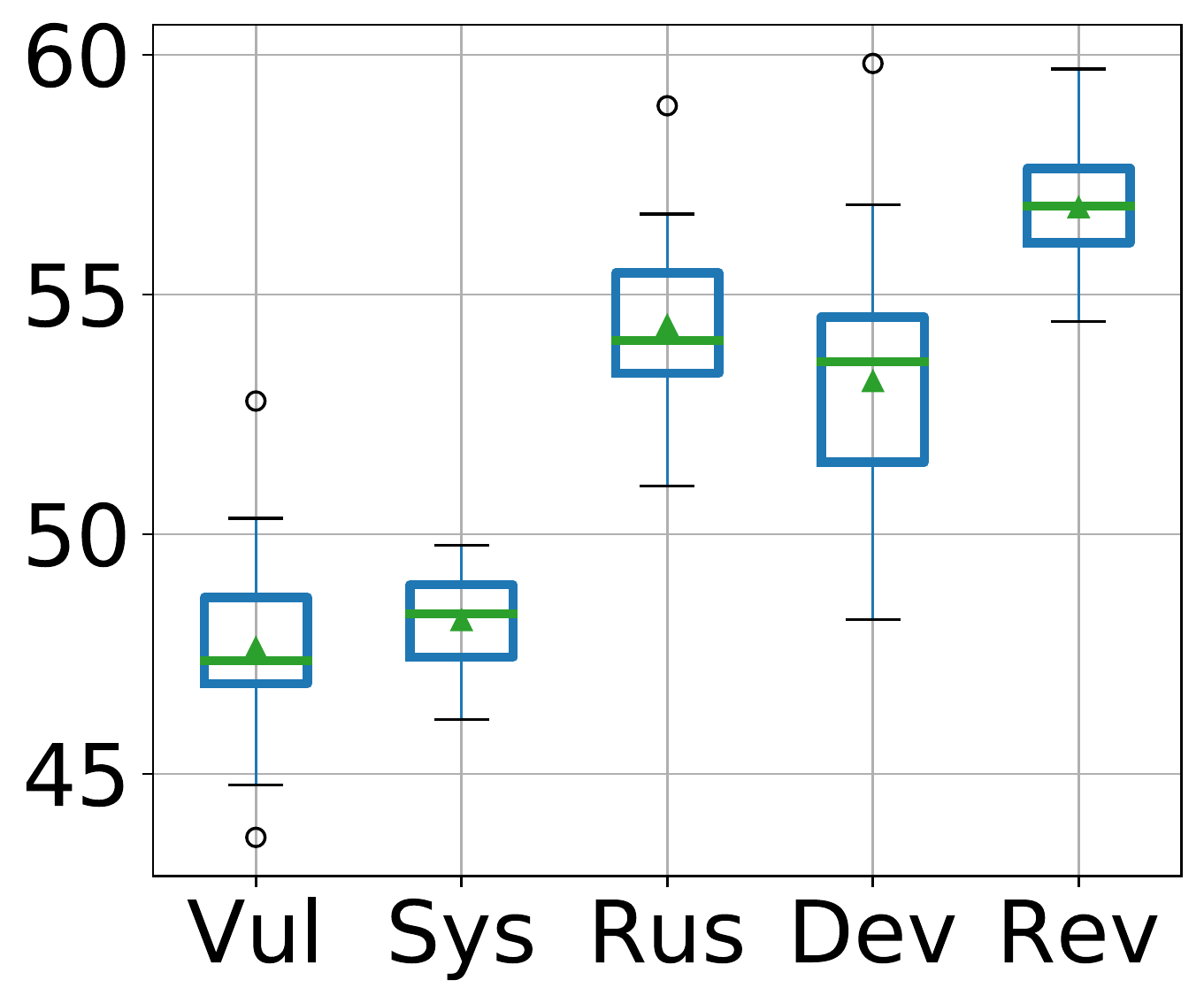}
        \caption{\small Precision}
        \label{fig:devign_pr_bp}
        \end{subfigure}
        \begin{subfigure}{0.30\linewidth}
        \includegraphics[width=\linewidth]{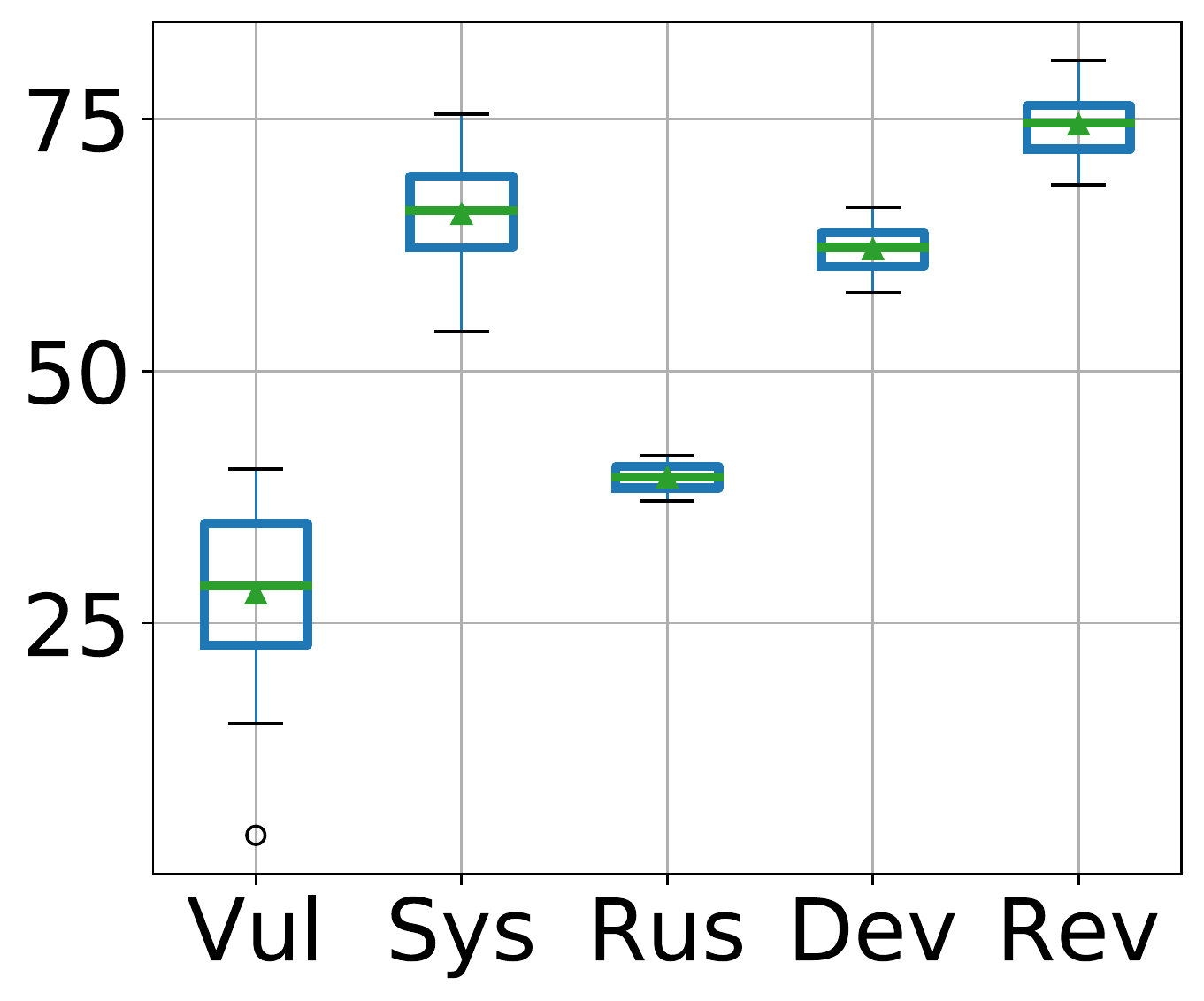}
        \caption{\small Recall}
        \label{fig:devign_rc_bp}
        \end{subfigure}
        \begin{subfigure}{0.30\linewidth}
        \includegraphics[width=\linewidth]{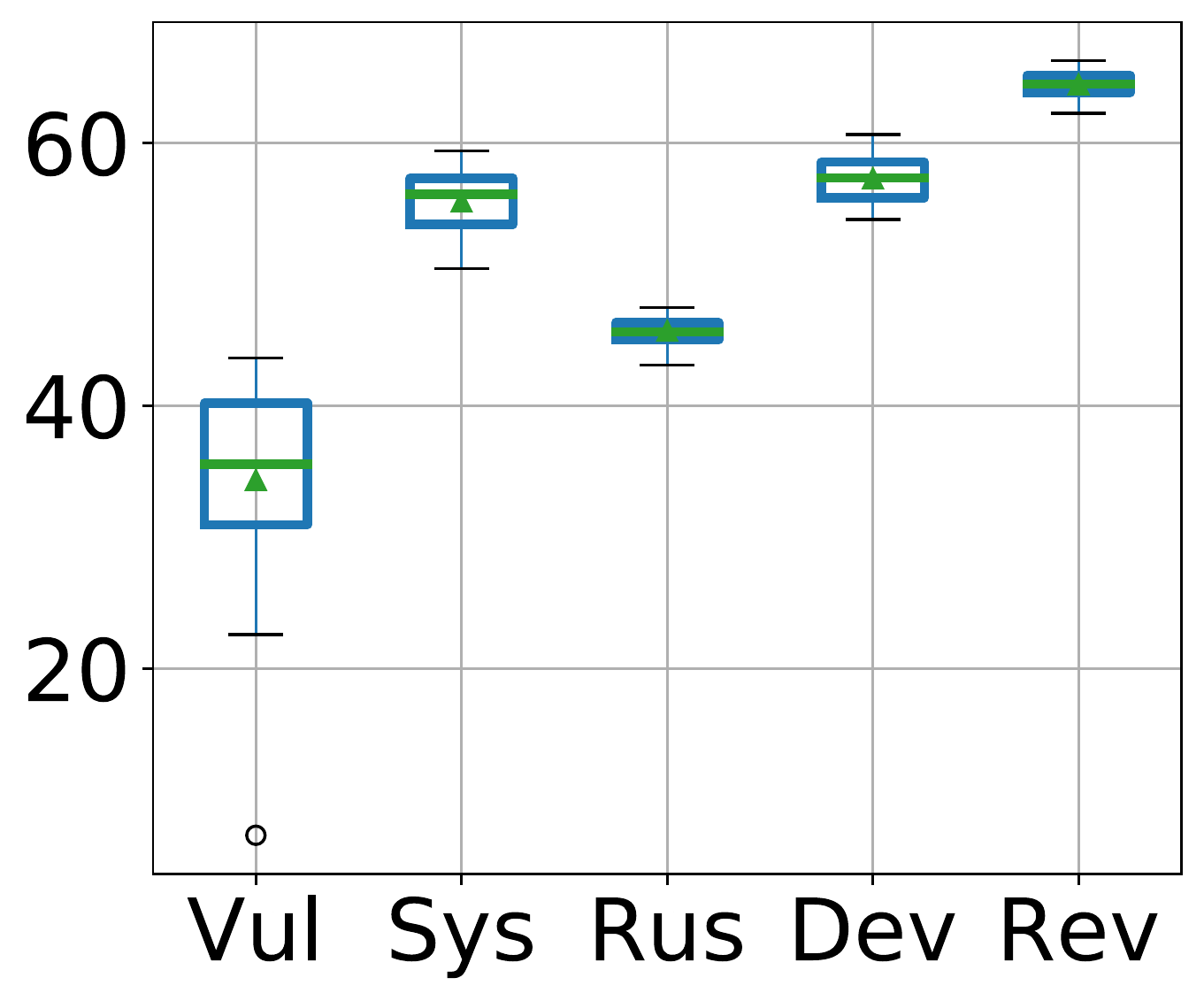}
        \caption{\small F1-score}
        \label{fig:devign_f1_bp}
        \end{subfigure}
        \caption{\small {Performance spectrum of \devigndata.}}
        \label{fig:devign_bp}
    \end{minipage}
    {\scriptsize Legends:~~~~Vul=VulDeePecker~\cite{li2018vuldeepecker},~~~~Sys=SySeVR~\cite{li2018sysevr},~~~~Rus=Russell~\etal~\cite{russell2018automated},~~~~Dev=Devign~\cite{zhou2019devign},~~~~Rev=\tool.}
\end{figure*}

\begin{figure}[!tb]
    \centering
    \begin{subfigure}{0.37\linewidth}
        \centering
        \includegraphics[width=\linewidth]{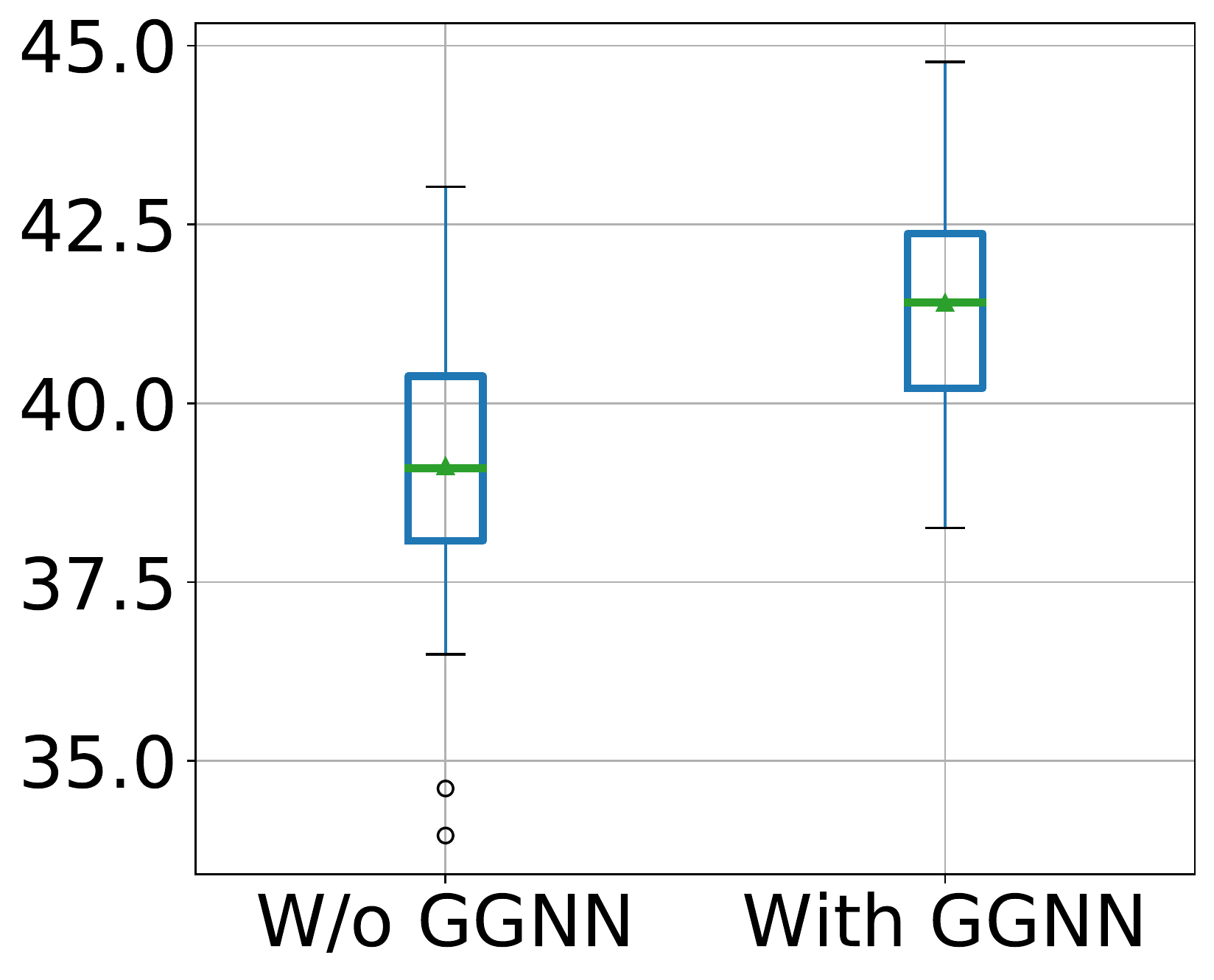}
        \caption{\realdata}
        \label{fig:verum_ggnn_abl}
    \end{subfigure}
    \begin{subfigure}{0.37\linewidth}
        \centering
        \includegraphics[width=\linewidth]{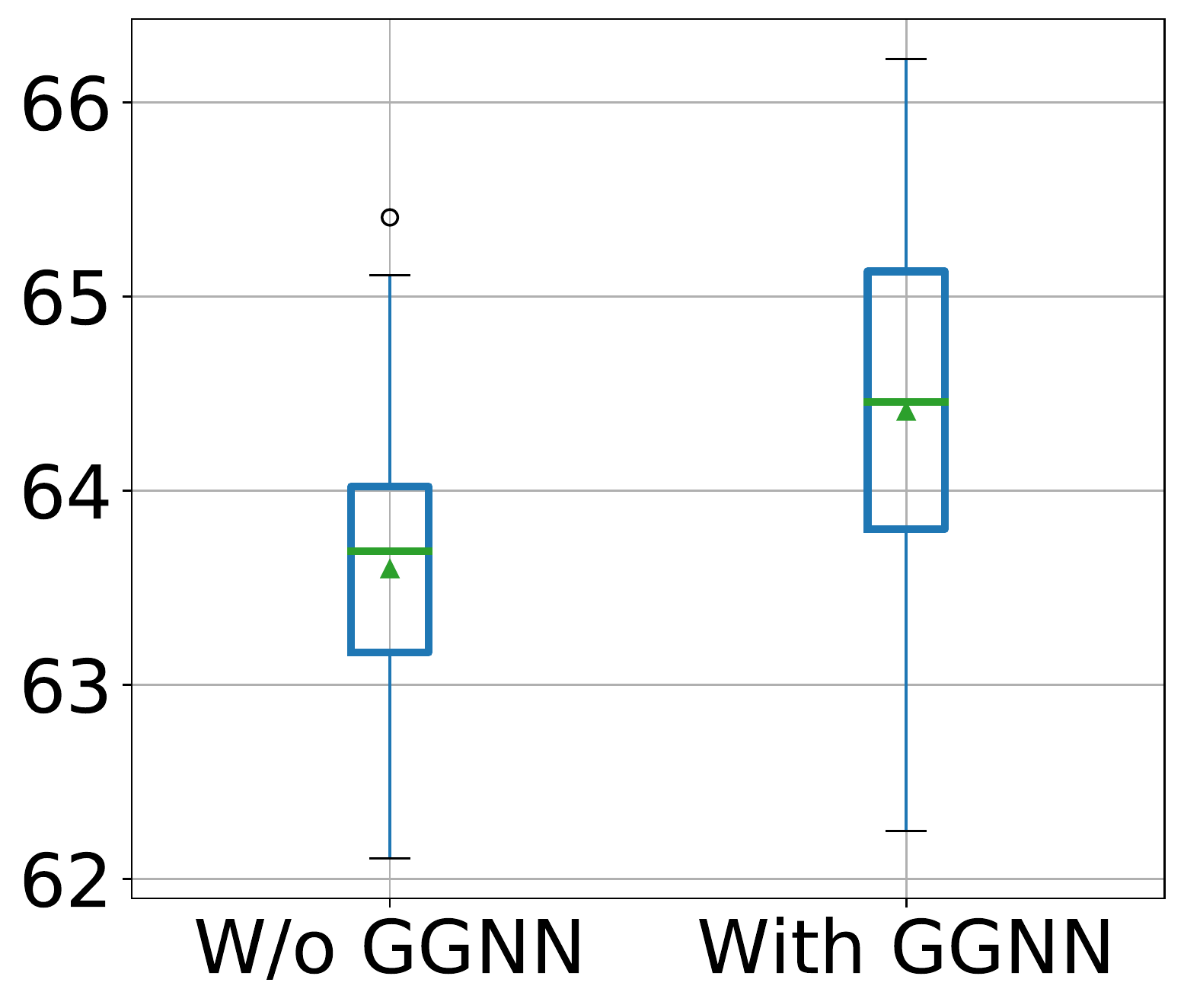}
        \caption{\devigndata}
        \label{fig:devign_ggnn_abl}
    \end{subfigure}
    \caption{\small{{Effect of GGNN in \tool's F1 score.} The performance increase in both datasets when node information is propagated to the neighboring node through GGNN. The effect size is 0.81 (large) for \realdata and 0.73 for \devigndata.}}
    \label{fig:ggnn_abl}
\end{figure}
\begin{figure}[!t]
    \centering
    \begin{subfigure}{0.37\linewidth}
        \centering
        \includegraphics[width=\linewidth]{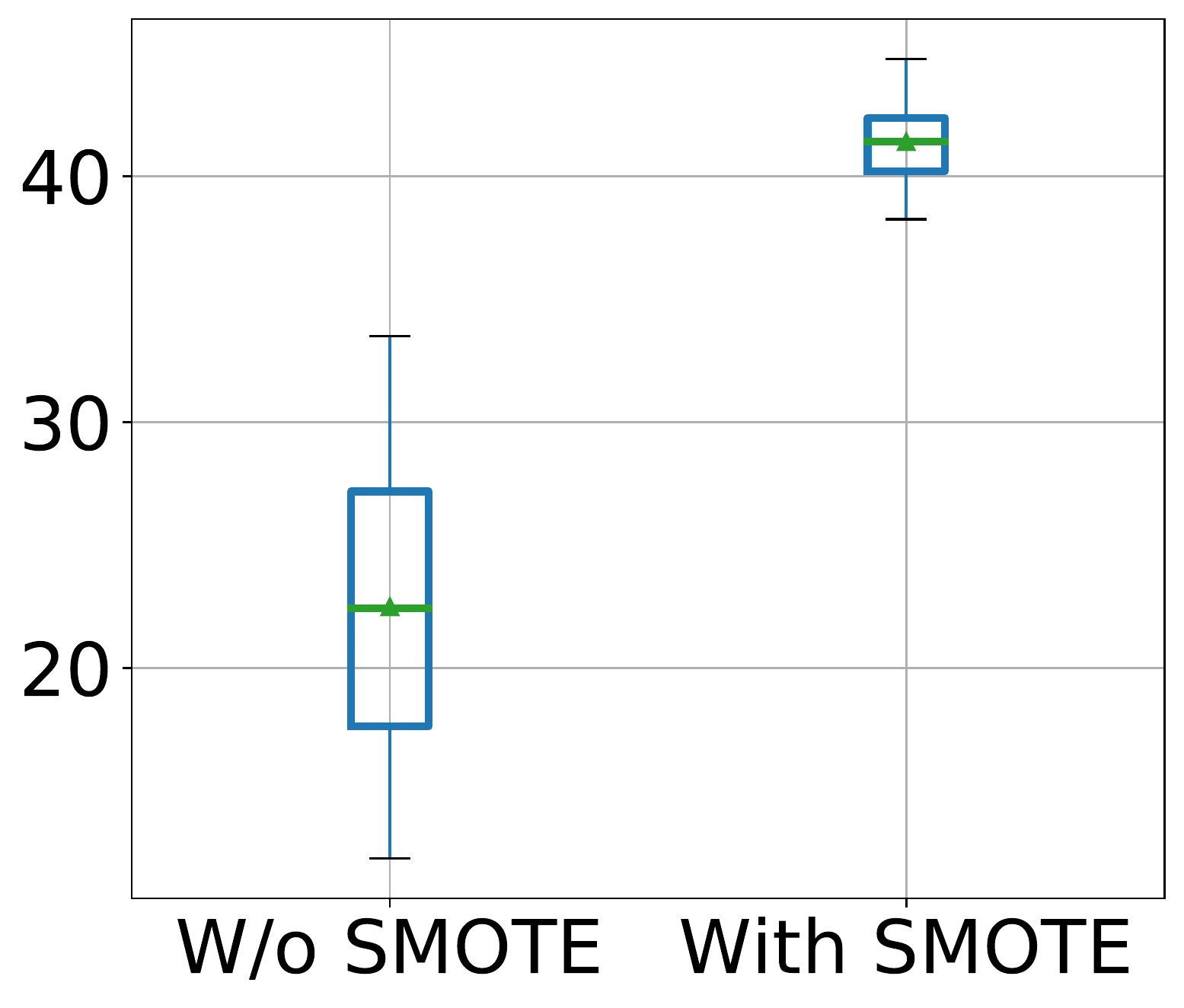}
        \caption{\realdata}
        \label{fig:verum_smote_abl}
    \end{subfigure}
    \begin{subfigure}{0.37\linewidth}
        \centering
        \includegraphics[width=\linewidth]{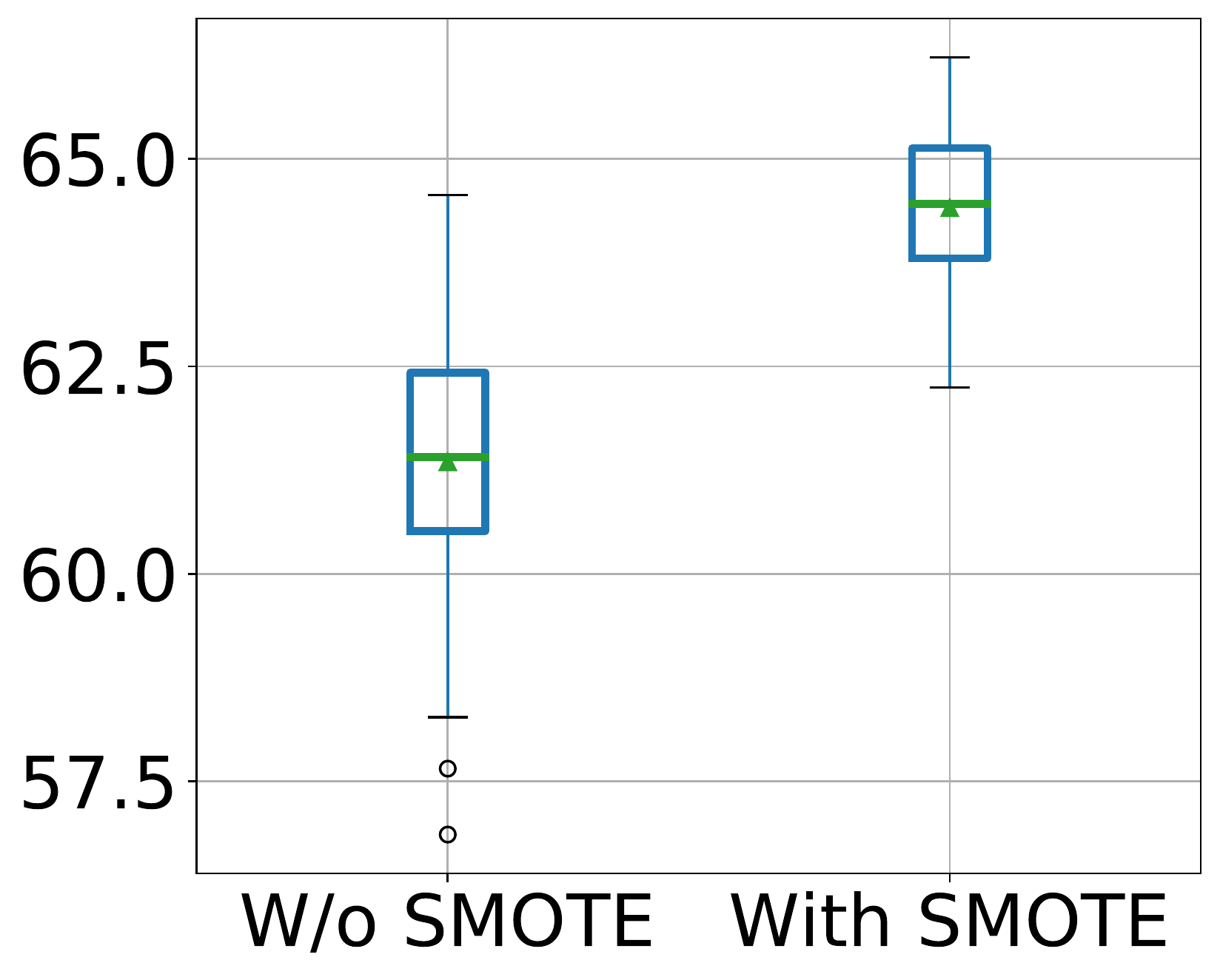}
        \caption{\devigndata}
        \label{fig:devign_smote_abl}
    \end{subfigure}
    \caption{\small {Effect of training data re-balancing in \tool's performance (F1-score).} In both datasets, re-balancing improves the performance of \tool. }
    \label{fig:balance_abl}
\end{figure}
\begin{figure}[!t]
    \centering
    
    \begin{subfigure}{0.37\linewidth}
        \centering
        \includegraphics[width=\linewidth]{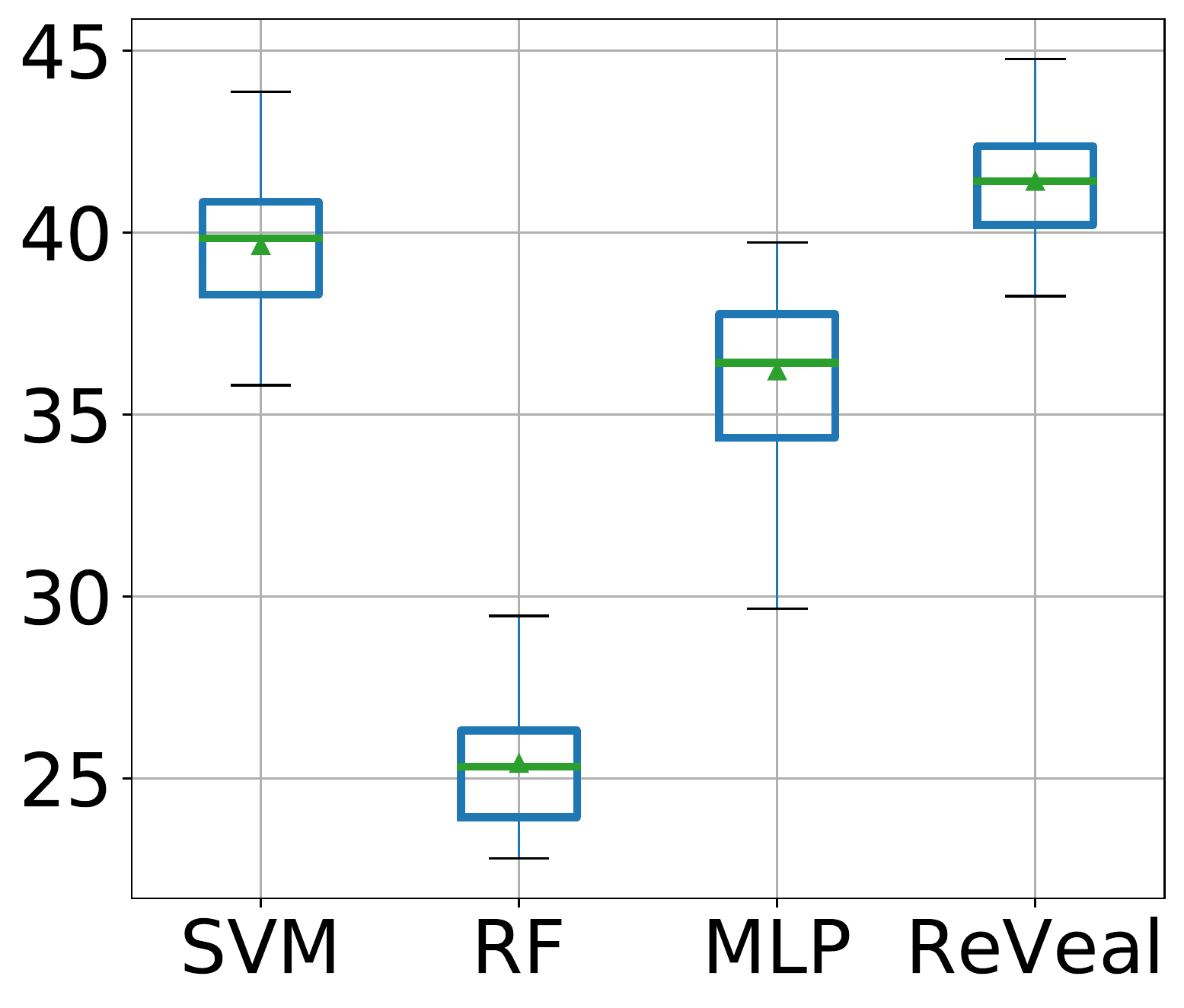}
        \caption{\realdata}
        \label{fig:verum_model_abl}
    \end{subfigure}
    \begin{subfigure}{0.37\linewidth}
        \centering
        \includegraphics[width=\linewidth]{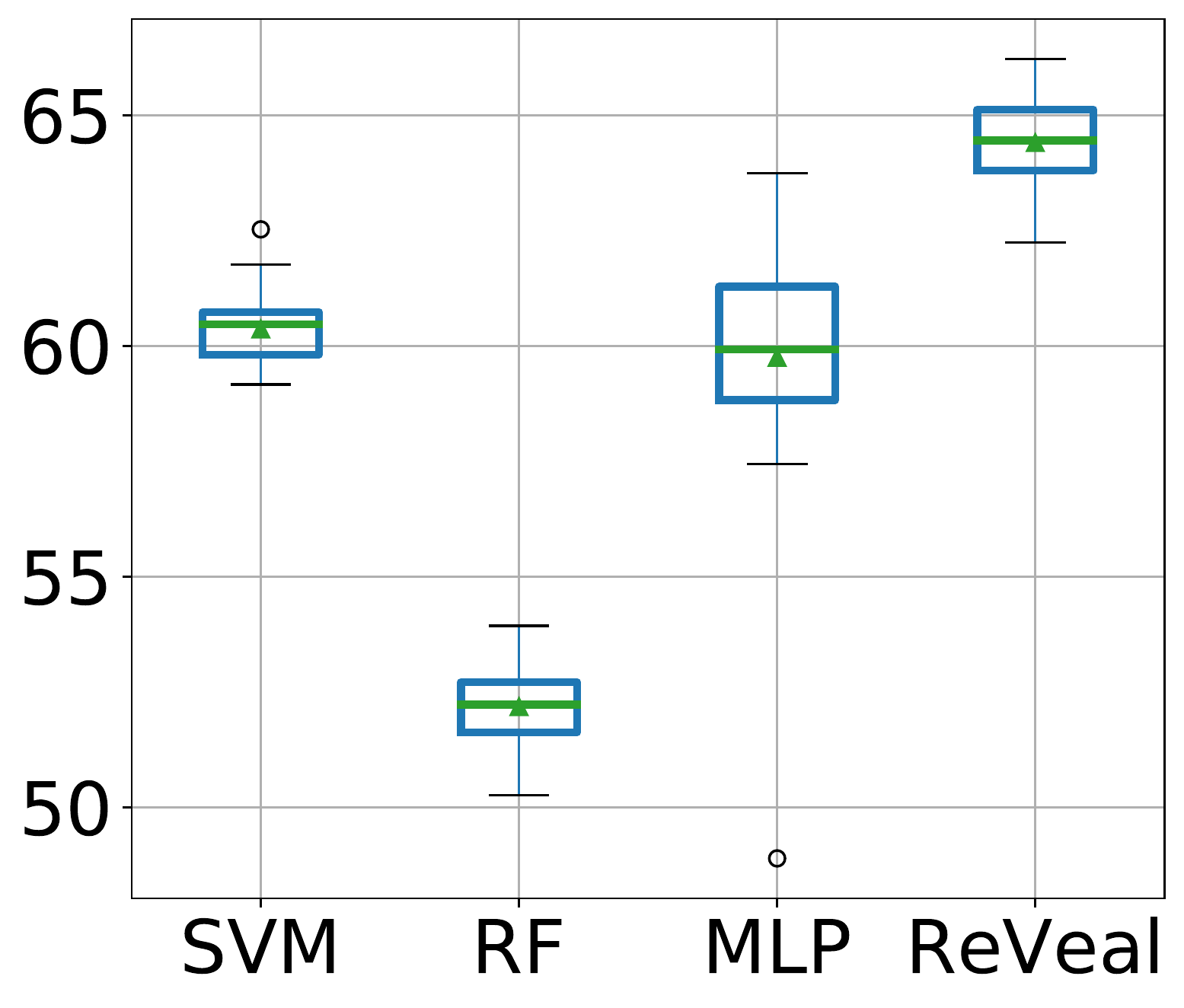}
        \caption{\devigndata}
        \label{fig:devign_models_abl}
    \end{subfigure}
    \caption{{\small \tool's performance (F1-score) in comparison to other machine learning models.}}
    \label{fig:model_abl}
\end{figure}

\subsection{How to improve \dlvp approaches? (RQ3)}
\label{sect:rq3}

\motivation In RQ2, we highlighted a number of challenges that limit the performance of existing \dlvp on real-world datasets. To address these challenges, we offer \tool --- a roadmap to help avoid some of the common problems that current state-of-the-art \vp methods face when exposed to real-world datasets. 

\approach
A detailed description of \tool is presented in \tion{pipeline}. Briefly, it works as follows: (i) input code fragment is converted to a feature vector with the help of a code property graph and GGNN~(\tion{feature_extraction}); (ii) the feature vectors are re-sampled using SMOTE~(\tion{smote}) that addresses potential data imbalance; and finally, (iii) a multi layer perceptron based representation learner is trained to learn a representation of the feature vectors that maximally separates the positive and negative classes~(\tion{representation_learning}). This pipeline offers the following benefits over the current state-of-the-art:
\be
\item \textit{Addressing duplication:} \tool does not suffer from data duplication. During pre-processing, input samples are converted to their corresponding code property graphs whose vertices are embedded with a GGNN and aggregated with an aggregation function. This pre-processing approach tends to create a unique feature for every input samples. So long as the inputs are not exactly the same, the feature vector will also not be the same. 
\item \textit{Addressing data imbalance:} \tool makes use of synthetic minority oversampling technique (SMOTE) to re-balance the distribution of vulnerable and non-vulnerable examples in the training data. This ensures that the trained model would be distribution agnostic and, therefore, better suited for real-world vulnerability prediction where the distribution of vulnerable and non-vulnerable examples is unknown.

\item \textit{Addressing model choice:} \tool extracts semantic as well as syntactic information from the source code using code property graphs. Using GGNN, each vertex embedding is updated with the embeddings of all its neighboring vertices. This further increases the semantic richness of the embeddings. This represents a considerable improvement to the current token-based and slicing-based models. As shown in~\fig{lemna_ggnn}, \tool can accurately predict the vulnerability here.

\item \textit{Addressing the lack of separability:} As shown in \fig{russell-t-sne}--\ref{fig:devign-t-sne}, the vulnerability class is almost inseparable from the non-vulnerability class in the feature space. 
To address this problem, \tool uses a representation learner that automatically learns how to re-balance the input feature vectors 
such that the vulnerable and non-vulnerable classes are maximally separated~\cite{6472238}. This offers significant improvements over the current state-of-the-art as shown in~\fig{reveal-t-sne}. Compared to the other approaches of \fig{russell-t-sne}--\ref{fig:devign-t-sne}, \tool exhibits the 
highest separation between the vulnerable and non-vulnerable classes (roughly $85\times$ higher than other GGNN based \vp).
\ee
We compare performance of \tool with existing \vp approaches of two real-world datasets, \ie, \devigndata and \tool data.

\observations
\Cref{fig:verun_bp,fig:devign_bp} compare the performance of \tool tool with other approaches. 
We observe that \tool offers noticeable improvements in all the metrics:

\noindent$\circ$~ \textit{\realdata:}~\tool performs best in terms of F1-score and recall. The median recall is $60.91\%$ ( $20.8\%$ more than that of SySeVR, the next best model) and median F1-score is $41.25\%$ ($11.38\%$ more than SySeVR).
This represents a $51.85\%$ and $36.36\%$ improvement in recall and F1 over SySeVR respectively. While \devign (another GGNN based \vp) produces a better precision, \devign's median recall $56.21\%$ less than that of \tool. This indicates that, compared to \devign, \tool can find larger number of true-positive vulnerabilities (resulting in a better recall) at the cost slightly more false-positives (resulting in a slightly lower precision). Overall, \tool's median F1-score is $11.38\%$ more than \devign, \ie, a $38.09\%$ improvement.

\noindent$\circ$~ \textit{\devigndata:} \tool outperforms other approaches in all performance metrics. \tool's median accuracy, precision, recall, and F1-scores are 5.01\%, 5.19\%, 13.11\%, and 12.64\% higher respectively than the next best approach.

In the rest of this research question, we investigate contribution of each component of \tool. Specifically, we study what improvements are offered by the use of (a) Graph neural network~(\tion{rq3c}); (b)~re-balancing training data with ~(\tion{rq3b}); and finally (c) representation learning~(\tion{rq3d}).

\subsubsection{Contribution of Graph Neural Network}
\label{sect:rq3c}
\begin{table}[!tpb]
    \centering
    \footnotesize
    \caption{\small Impact of GGNN in \tool's performance [Median (IQR)].}
    \resizebox{0.9\linewidth}{!}{
        \begin{tabular}{@{}l|r|c|c|c|c}
            \hlineB{2}
            \textbf{Dataset} & \textbf{Approach} &  \textbf{Accuracy} & \textbf{Precision}  & \textbf{Recall} & \textbf{F1-score}\bigstrut\\
            \hlineB{2}
            \multirow{4}{*}{\rotatebox{90}{\tool}~~\rotatebox{90}{~dataset}} & 
            \tool  & 83.69 & 29.48 & 57.69 & 39.09\bigstrut[t]\\	
            & w/o GGNN & \textit{(1.60)} & \textit{(2.69)} & \textit{(7.85)} & \textit{(2.30)}\bigstrut[b]\\
            \cline{2-6}
            & 
            \tool &  \textbf{84.37} & 30.91 & \textbf{60.91} & \textbf{41.25} \bigstrut[t]\\
            & with GGNN & \textit{(\textbf{1.73})} & \textit{(2.76)} & \textit{(\textbf{7.89})} & \textit{(\textbf{2.28})}\bigstrut[b]\\ 
           \hlineB{2}
           
            \multirow{4}{*}{\rotatebox{90}{FFMpeg+}~~\rotatebox{90}{~Qemu}} & 
            \tool  & 53.87 & 49.60  & \textbf{89.25 } & 63.69  \bigstrut[t]\\
            & w/o GGNN & \textit{(2.69)} & \textit{(1.68)} & \textit{(\textbf{3.78})} & \textit{(0.85)} \bigstrut[b]\\
            \cline{2-6}
            & 
           \tool & \textbf{62.51} & \textbf{56.85} & 74.61 & \textbf{64.42} \bigstrut[t]\\
           & with GGNN & \textit{(\textbf{0.90})} & \textit{(\textbf{1.54})} & \textit{(4.31)} & \textit{(\textbf{1.33})} \bigstrut[b]\\
           \hlineB{2}
        \end{tabular}
    }
    \label{tab:comparison_wo_ggnn}
\end{table}

To understand the contribution of GGNN, we create a variant of \tool without GGNN. In this setup, we bypass the use GGNN and aggregate the initial vertex features to create the graph features. Further, we create another variant of \tool that uses \textit{only GGNN} \textit{without} re-sampling or representation learning. 

\Cref{fig:ggnn_abl} shows the F1-scores for the above setup. We observe that, in both \realdata and \devigndata, F1-score increases when we use GGNN in \tool's pipeline. We observe that the improvements offered by the use of GGNN is statistically significant (with a p-value of 0.0002 in \realdata, and 0.001 in \devigndata). Further, when we perform the A12 effect size~\cite{hess2004robust} with 30 independent experiment runs in each case, we found that the the effect size is 81\% for \realdata and 73\% for \devigndata. This means that 81\% of the times \tool performs better with GGNN than it does without GGNN in \realdata and 73\% in \devigndata. Both of those effect sizes are considered large indicating \tool with GGNN's f1-score distribution is better than \tool without GGNN. 

We contend that, since GGNN embeds the neighbors' information in every vertex, vertices have richer information about the graph. Thus \tool's classification model have more information at its disposal to reason about. The result indicates that when vertices assimilate information from neighboring vertices, \vp performance increases. 

\subsubsection{Effect of Training Data Balancing}
\label{sect:rq3b}

\begin{table}[!tpb]
    \centering
    \footnotesize
    \caption{\small Impact of re-balancing training data in \tool's performance (Median/IQR). \devigndata is almost balanced, thus further re-balancing do not impact performance that much. However, in \realdata, training data re-balancing improved the performance significantly.}
    \resizebox{0.9\linewidth}{!}{
        \begin{tabular}{@{}l|r|c|c|c|c}
            \hlineB{2}
            \textbf{Dataset} & \textbf{Approach} &  \textbf{Accuracy} & \textbf{Precision}  & \textbf{Recall} & \textbf{F1-score}\bigstrut\\
            \hlineB{2}
            \multirow{4}{*}{\rotatebox{90}{\tool}~~\rotatebox{90}{~dataset}} & 
        \multirow{1}{*}{W/O} & 90.48 & \textbf{46.23}  & 15.09  & 22.44  \bigstrut[t]\\	
            & \multirow{1}{*}{Re-balance} & \textit{(0.93)} & \textit{(\textbf{11.30})} & \textit{(9.09)} & \textit{(9.56)} \bigstrut[b]\\
            \cline{2-6}
            & 
            \multirow{1}{*}{With} &  \textbf{84.37} & {30.91} & \textbf{60.91} & \textbf{41.25} \bigstrut[t]\\
            & Rebalance & \textit{(\textbf{1.73})} & \textit{({2.76})} & \textit{(\textbf{7.89})} & \textit{(\textbf{2.28})} \bigstrut[b]\\
           \hlineB{2}
           
            \multirow{4}{*}{\rotatebox{90}{FFMpeg+}~~\rotatebox{90}{~Qemu}} & 
            \multirow{1}{*}{W/O} & 62.94  & 58.86  & 64.20 & 61.40 \bigstrut[t]\\	
            & Re-balance & \textit{(1.40)} & \textit{(2.88)} & \textit{(6.76)} & \textit{(1.91)} \bigstrut[b]\\
            \cline{2-6}
            & 
            \multirow{2}{*}{With} & \textbf{62.51 } & \textbf{56.85 } & \textbf{74.61 } & \textbf{64.42 } \bigstrut[t]\\
            & Re-balance & \textit{(\textbf{0.90})} & \textit{(\textbf{1.54})} & \textit{(\textbf{4.31})} & \textit{(\textbf{1.33})} \bigstrut[b]\\
           \hlineB{2}
        \end{tabular}
    }
    \label{tab:comparison_wo_smote}
\end{table}

To understand the contribution of SMOTE, we deploy two variants of \tool one with SMOTE and one without. Note that, \tool uses SMOTE as an off-the shelf data balancing tool. Choice of which data-balancing tool should be used is a configurable parameter in \tool's pipeline. 

\Cref{fig:balance_abl}  illustrates the effect of using data re-sampling in \tool's pipeline. We observe that re-balancing training data improves \tool's performance in general. The more skewed the dataset, the larger the improvement. In \devigndata, non-vulnerable examples populates roughly $55\%$ of the data. There, using SMOTE offers only a $3\%$ improvement in F1-score (see~\fig{devign_smote_abl}). However, in \realdata, non-vulnerable examples populates $90\%$ of the data, there we obtain more than $22\%$ improvement in F1-score compared to not using SMOTE (see~\fig{verum_smote_abl}). 
Without SMOTE, the precision of \tool tool improves and reaches up to 46.23\% (see~\Cref{tab:comparison_wo_smote} in Appendix), highest achieved precision among all the experimental settings. However, this  setting suffers from low recall due to data imbalance. Thus, if an user cares more about precision over recall, SMOTE can be turned off, and vice versa.

\subsubsection{Effect of Representation Learning}
\label{sect:rq3d}


\begin{table}[!tpb]
    \centering
    \footnotesize
    \caption{\small \tool's performance in comparison to other baseline models -- \ie Random Forest (RF), Support Vector Machine (SVM), MLP with NLL Loss (MLP$^\dagger$). \tool achieves better performance that other max margin model (SVM), and \tool's loss function improves the performance with respect to the MLP$^\dagger$.}
    \resizebox{0.9\linewidth}{!}{
        \begin{tabular}{@{}l|r|c|c|c|c}
            \hlineB{2}
            \textbf{Dataset} & \textbf{Approach} &  \textbf{Accuracy} & \textbf{Precision}  & \textbf{Recall} & \textbf{F1-score}\bigstrut\\
            \hlineB{2}
            \multirow{8}{*}{\rotatebox{90}{\tool}~~\rotatebox{90}{~dataset}} & 
            \multirow{2}{*}{RF} & \textbf{85.46} & 24.15 & 26.58 & 25.32 \bigstrut[t]\\	
            & & \textit{(\textbf{0.62})} & \textit{(3.03)} & \textit{(2.08)} & \textit{(2.39)} \bigstrut[b]\\
            \cline{2-6}
            &
            \multirow{2}{*}{MLP$^\dagger$} & 84.81 & 29.35 & 47.51 & 36.42 \bigstrut[t]\\	
            & & \textit{(1.80)} & \textit{(3.51)} & \textit{(4.77)} & \textit{(3.40)} \bigstrut[b]\\
            \cline{2-6}
            &
            \multirow{2}{*}{SVM} & 82.61  & 29.42  & \textbf{61.20 } & 39.85  \bigstrut[t]\\	
            & & \textit{(0.43)} & \textit{(2.45)} & \textit{(\textbf{1.86})} & \textit{(2.54)} \bigstrut[b]\\
            \cline{2-6}
            &
            \multirow{2}{*}{\tool} &  84.37 & \textbf{30.91} & 60.91 & \textbf{41.25} \bigstrut[t]\\
            & & \textit{(1.73)} & \textit{(\textbf{2.76})} & \textit{(7.89)} & \textit{(\textbf{2.28})} \bigstrut[b]\\
           \hlineB{2}
           
            \multirow{8}{*}{\rotatebox{90}{FFMpeg+}~~\rotatebox{90}{~Qemu}} & 
            \multirow{2}{*}{RF} & 57.34  & 53.62  & 50.90  & 52.23  \bigstrut[t]\\	
            & & \textit{(0.84)} & \textit{(1.50)} & \textit{(1.23)} & \textit{(1.09)} \bigstrut[b]\\
            \cline{2-6}
            &
            \multirow{2}{*}{MLP$^\dagger$} & 61.43 & 56.87 & 63.36 & 59.93 \bigstrut[t]\\	
            & & \textit{(1.38)} & \textit{(2.04)} & \textit{(4.81)} & \textit{(2.46)} \bigstrut[b]\\
            \cline{2-6}
            & 
            \multirow{2}{*}{SVM} & 61.84 & 57.66 & 63.26  & 60.47  \bigstrut[t]\\	
            & & \textit{(0.55)} & \textit{(1.18)} & \textit{(1.55)} &\textit{ (0.92) }\bigstrut[b]\\
            \cline{2-6}
            &
            \multirow{2}{*}{\tool} & 62.51 & 56.85  & \textbf{74.61} & \textbf{64.42} \bigstrut[t]\\
            & & \textit{(0.90)} & \textit{(1.54)} & \textit{(\textbf{4.31})} & \textit{(\textbf{1.33})} \bigstrut[b]\\
           \hlineB{2}
        \end{tabular}
    }
    \label{tab:comparison_models}
\end{table}

In order to understand the contribution of representation learning, we replace representation learning with three other learners: (a) Random Forest (a popular decision tree based classifier used by other \vp approaches like Russell~\etal~\cite{russell2018automated}); (b) SVM with an RBF kernel which also attempts to maximize the margin between vulnerable and non-vulnerable instances~\cite{baesens2000empirical}; and (c) An off-the-shelf Multi-Layer Perceptron which uses a log-Likelihood loss~\cite{platt1999probabilistic}. 

\Cref{fig:model_abl} shows the \tool's performance with different classification models. In both \realdata and \devigndata, our representation learner with triplet loss achieves the best performance. \tool's median F1-score is 62.8\%, 13.3\%, 3.5\% higher than that of RF, MLP, and SVM baselines respectively in \realdata. For \devigndata improvement in median F1-score is 23.33\%, 7.5\%, 6.5\% over RF, MLP, and SVM respectively.

Max-margin models results in better performance in classifying vulnerable code in general. \tool with the representation learner performs statistically and significantly better than SVM in both \realdata and \devigndata (with $\textit{p-values}<0.01$ and $\textit{A12}>0.6$). This is likely because SVM is a shallower than a representation learning model that propagates losses across several perceptron layers.


\begin{result}
The performance of \dlvp approaches can be significantly improved using the \tool pipeline. The use of GGNN based feature embedding along with SMOTE and representation learning remedies data-duplication, data imbalance, and lack of separability. 
\tool produces improvements of 
up to 33.57\% in precision and 128.38\% in recall over state-of-the-art methods.
\end{result}
\section{Discussion}
\label{sec:discussion}

\subsection{Vulnerability Detection in Real World}


The usefulness of a source code vulnerability detection tool depends on its use case scenario. Ideally, in a real-world scenario, developers would deploy a trained vulnerability prediction model to identify vulnerable functions from a codebase. In simple terms, given all the functions in the code base, developers would want to locate the vulnerable function. As discussed in \Cref{sect:background}, evaluating such a scenario is paramount in understanding the usefulness of an approach. Existing approaches show very little to no evaluation of how respective approaches perform in such a real-world scenario. For instance, Devign~\cite{zhou2019devign} showed a simulated real imbalanced evaluation (we refer to Table 3 in Devign’s original paper). However, for that simulation, they randomly sampled their version of the test data to contain 10\% vulnerable example. Their reported results also show the drop in performance in real world imbalanced settings. Nevertheless, we hypothesize that their evaluation method does not truly reflect how a method will perform in the real world since the imbalance in their dataset is artificial. In this study, we propose a method to simulate such an evaluation scenario. Therefore, we hope such evaluation settings help drive new research in vulnerability detection.

\subsection{Vulnerability Data and Tangled commits}

\begin{figure}[!htb]
    \centering
    \includegraphics[width=0.6\linewidth]{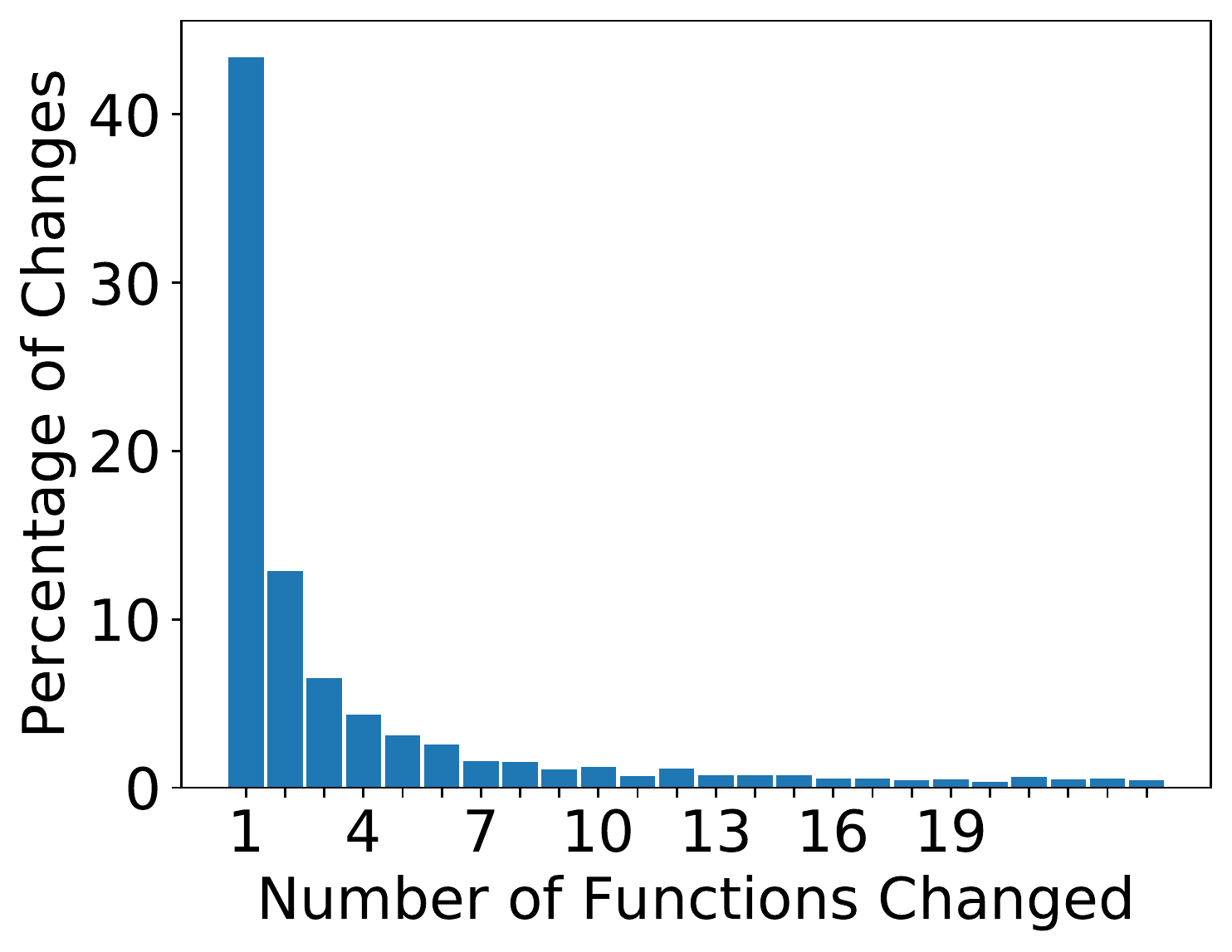}
    \caption{\small Histogram of number of changed function in Vulnerability fix patches.}
    \label{fig:vulnerability_stat}
\end{figure}
Tangled commits have long been studied in software engineering~\cite{dias2015untangling, sothornprapakorn2018visualizing, wang2019cora, tao2015partitioning} and a major setback for software evolution history driven research~\cite{herzig2013impact, herzig2016impact, kirinuki2014hey}. Developers often combine more than one unrelated or weakly related changes in code in one commit~\cite{herzig2013impact} causing such a commit to be entanglement of more than one changes. 
Our collected \tool data is also subject to such a threat of containing tangled code changes. Thus, we investigate the characteristics of number of function changes in vulnerability-fix patches. \Cref{fig:vulnerability_stat} shows a histogram of number of functions that are changed per Vulnerability-fix patches. Most of the patches change very small number of functions. 80\% of the changes account for 12 or fewer number of function changes. 

To validate that the empirical finding in the paper are not biased by the tangled commits, we created an alternate version of \tool data, where we removed any patch that changes more than one function from consideration. In that version of \tool data, we find that \tool achieves 26.33\% f1 score. In contrast, if we do not use representation learning, \tool's f1 score drops to 22.95\%. If we do not use the data balancing, \tool's performance drops to 13.13\%. When we remove GGNN from \tool's pipeline, f1 score drops to 22.82\%. These results corroborates the importance of GGNN, data balancing and representation learning in \tool's pipeline irrespective of existence of tangled code changes.

\section{Related Work}
\label{sec:rel}

\noindent
There have been a wide array of ML-based \vp research~\cite{lin2017poster, ban2019performance, meng2016automatically,joh2008vulnerability, jie2016survey}. 
Yamaguchi \etal~\cite{yamaguchi2013chucky} applied anomaly detection techniques on embeddings produced from static tainting to discover missing conditions such as input validation. Perl \etal~\cite{perl2015vccfinder} used
commit messages to detect the vulnerabilities of a program. These work leveraged Support Vector Machine (SVM) for VP. Li \etal~\cite{li2016vulpecker} used multi-class SVM to detect different class of vulnerabilities. 
Recently, \dlvp has been subject to much research in both static-~\cite{russell2018automated} and 
dynamic~\cite{she2019neuzz}-analysis scenario. 
However, static settings are more popular~\cite{larochelle2001statically} using code slices~\cite{li2018sysevr, li2018vuldeepecker}, trees~\cite{mou2014TBCNN}, graphs~\cite{zhou2019devign} etc. Although most prominent approaches~\cite{li2018vuldeepecker, li2018sysevr, russell2018automated, zou2019muvuldeepecker} use token based representation of code, recent graph based modeling showed success in \vp~\cite{zhou2019devign}.



 \textit{Code Property Graph (CPG)}, introduced by Yamaguchi~\etal~\cite{yamaguchi2014modeling}, models the combined semantic and syntactic information of a program. The CPG is a joint data structure that leverages the information from abstract syntax trees, control flow graphs and program dependency graphs. CPG has shown to be robust in reasoning about vulnerabilities~\cite{yamaguchi2015automatic, yamaguchi2014modeling, zhou2019devign}. Thus, \tool uses CPG to extract graph based features. In addition, \tool reduces data imbalance bias (through re-sampling in feature space) and learns to maximize separation between vulnerable and non-vulnerable examples.
 
 There are several choices of techniques for reducing data imbalance~\cite{lusa2013smote, chawla2002smote, barua2012mwmote}, all of which use different strategies for balancing any imbalanced datasets. We choose SMOTE in \tool's pipeline as it has shown to be successful in other software engineering related tasks~\cite{agrawal2018better, alsawalqah2017hybrid, guo2019identify, pak2018empirical, pears2014synthetic}. 

\section{Conclusion}

In this paper, we systematically study different aspects of Deep Learning based Vulnerability Detection to effectively find real world vulnerabilities. We empirically show different shortcomings of existing datasets and models that potentially limits the usability of those techniques in practice. Our investigation found that existing datasets are too simple to represent real world vulnerabilities and existing modeling techniques do not completely address code semantics and data imbalance in vulnerability detection. Following these empirical findings, we propose a framework for collecting real world vulnerability dataset. We propose \tool as a configurable \vp tool that addresses the concerns we discovered in existing systems and demonstrate its potential towards a better \vp tool. 

\balance
\section*{Acknowledgements}
We would like to thank Yufan Zhuang for initial help in data collection. We also thank Dr. Suman Jana for their extensive feedback on this paper.

\bibliographystyle{IEEEtran}
\bibliography{references}


\end{document}